\documentclass[useAMS, usenatbib, usegraphicx]{mn2e}
\usepackage{epsfig}
\usepackage{color}

\def\imp{$\frac{\rm b}{\rm r_{\rm cl}}$}
\def\velp{$\frac{v_{\rm 0}}{v_{\rm b}}$}

\title[Black Hole Cloud Collisions]{Simulations of Direct Collisions of Gas Clouds with the Central Black Hole}

\author[C. Alig, A. Burkert, P.H. Johansson, M. Schartmann]{C. Alig$^{1,2}$, A. Burkert$^{1,2,3}$, P.H. Johansson$^{1}$, M. Schartmann$^{1,2}$\\
$^{1}$Universit\"ats-Sternwarte M\"unchen, Scheinerstr.1, D-81679 M\"unchen, Germany \\
$^{2}$ Max-Planck-Institut f\"ur extraterrestrische Physik, Giessenbachstr., D-85748 Garching, Germany \\
$^{3}$ Max-Planck-Fellow}

\begin{document}

\date{Accepted 2010 October 25. Received 2010 October 25; in original form 2010 August 7}
\pagerange{\pageref{firstpage}--\pageref{lastpage}} \pubyear{2010}
\maketitle

\label{firstpage}
\begin{abstract}

We perform numerical simulations of clouds in the Galactic Centre (GC) engulfing the nuclear super-massive black hole
and show that this mechanism leads to the formation of gaseous accretion discs with properties
that are similar to the expected gaseous progenitor discs that fragmented into the observed
stellar disc in the GC.
As soon as the cloud hits the black hole, gas with opposite angular momentum relative to the black hole collides downstream.
This process leads to redistribution of angular momentum and dissipation of kinetic energy,
resulting in a compact gaseous accretion disc.
A parameter study using thirteen high resolution simulations of homogeneous
clouds falling onto the black hole and engulfing
it in parts demonstrates that this mechanism is able to produce gaseous accretion discs that could potentially 
be the progenitor of the observed stellar disc in the GC.
A comparison of simulations with different equations of state (adiabatic, isothermal and full cooling)
demonstrates the importance of including a detailed thermodynamical description. However the simple
isothermal approach already yields good results on the radial mass transfer and accretion rates,
as well as disc eccentricities and sizes. We find that the cloud impact parameter strongly
influences the accretion rate whereas the impact velocity has a small affect on the accretion rate.

\end{abstract}



\section{Introduction}
\label{intro}

Observations reveal an interesting feature in our Galactic Centre (GC):
one \citep{2009ApJ...690.1463L} or two \citep{2003ApJ...594..812G, 2005ApJ...620..744G, 2006ApJ...643.1011P,
2009ApJ...697.1741B} sub-parsec-scale rings of young stars near the radio source SgrA* 
which cannot be explained by normal means of star formation due to the hostility of the environment.
Tidal forces would disrupt typical molecular clouds in the vicinity of the central black hole, preventing
their normal condensation into stars due to the self-gravity induced collapse of the cloud.

Observations find around 100 young and massive stars distributed in a warped clockwise rotating disc/ring and a second
inclined counter-clockwise rotating disc/ring, the existence of which is still a matter of debate
\citep{2009ApJ...690.1463L}.
The observed outer edge of the system is at around 0.5 pc. This estimate is still increasing
due to the addition of newly observed stars
belonging to the disc structure. The inner edge is at around 0.04 pc \citep{2003ApJ...594..812G}.
The mean eccentricity of the clockwise rotating system is measured to be 0.34 $\pm$ 0.06 \citep{2009ApJ...697.1741B}.

The currently most plausible formation scenario postulates that the stellar rings formed by fragmentation
of self-gravitating (in some cases eccentric) accretion discs
(\citealt{1977SvAL....3..134P, 1980SvAL....6..357K, 1987Natur.329..810S,
2005MNRAS.364L..56R, 2008ApJ...674..927A} and first applied to our GC in \citealt{1998MNRAS.294...35S,
2003ApJ...590L..33L}).
These studies concentrated mainly on the fragmentation of an already existing accretion disc. The
question how these discs formed in the first place has only recently  been discussed, e.g. by
\citet{2008ApJ...683L..37W}, \citet{2008arXiv0805.0185M}, \citet{2009MNRAS.394..191H} and \citet{2008Sci...321.1060B}.
In their models, reviving the origin of the stars from a molecular cloud, the accretion discs are built up
by the rapid deposition of gas around the central black hole through the infall and tidal disruption
of a gas cloud. However, here the problem arises that a gas cloud that most likely formed far from the GC where
tidal forces are inefficient must be placed on an orbit with a close passage around the black hole.

Furthermore, the cloud should not be disrupted by internal star formation processes
before encountering the black hole.  \citet{2008arXiv0805.0185M}, \citet{2009MNRAS.394..191H}
and \citet{2008Sci...321.1060B} studied the formation of a disc by a cloud or clouds
passing the central black hole at some distance larger than the cloud radius.
However, \citet{2008ApJ...683L..37W} showed that it is very unlikely for a cloud to be captured by the black hole
without engulfing it during the encounter, as only a small set of initial parameters would lead to such an
event. Thus \citet{2008ApJ...683L..37W} proposed a model in which the infalling cloud covers the black hole in
parts during its passage. This process efficiently redistributes angular momentum through the collision
of material streaming around the black hole with opposite angular momentum, finally resulting in a
compact accretion disc. 
A related scenario for the effective cancellation of angular momentum has also
been simulated by \citealt{2009MNRAS.394..191H}, who studied the collision of two 
clouds with opposite angular momentum close to the GC black hole.

In this first paper we elaborate on this scenario and study direct encounters of gas clouds with the central black
hole in detail using high-resolution numerical simulations. 
We focus on the resulting gaseous disc properties for cloud impact parameters that 
match the Milky Way GC. The resulting gas disc could then be the progenitor of the observed stellar disc.
Clearly, a potential caveat is the fact that
the GC stellar disc is already many orbital periods old and could have undergone substantial secular
dynamical evolution (see e.g. \citealt{2009MNRAS.398..535U, 2009MNRAS.398..429L, 2009ApJ...697L..44M}).
Arguments have however also been presented
that the observed stellar disc properties are more or less set by the initial gaseous
disc properties (\citealt{2007ApJ...654..907A, 2008MNRAS.388L..64C}).
The estimated mass of the gaseous disc that formed the stars
is expected to be around 10$^{4-5}$ M$_{\sun}$ \citep{2005A&A...437..437N, 2006MNRAS.366.1410N}.
Some fraction of the disk might have been accreted onto the central
black hole. Note, that here, we are not focusing on this long term viscous evolution of the resulting
accretion disc which we are in any case unable to follow and resolve with our current numerical scheme.
Instead we focus on the formation of the initial gaseous disc resulting
from the collision of the cloud with the black hole.
In a follow-up paper we will discuss fragmentation and star formation inside the accretion disc.

This paper is structured as follows. In Section \ref{motiv} we present a motivation for our model.
In Section \ref{num} the numerical method and the initial setup of the model are presented.
The main results as a function of varying impact parameters, initial cloud velocities and different equations
of state are presented in Section \ref{results}. 
Finally, we summarise and discuss our findings in Section \ref{summary}.
In the appendix, code tests to demonstrate the numerical stability of our results are presented.


\section{Motivation}
\label{motiv}

\begin{figure}
\begin{center}
\includegraphics[width=9cm]{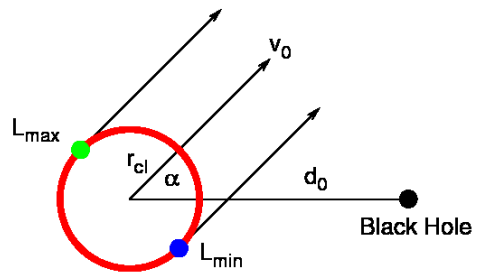}
\end{center}
\caption{Coordinate system used to calculate the cloud angular momentum. 
We define d$_{\rm 0}$ to be the distance from the black hole to the cloud, v$_{\rm 0}$ to be the initial
absolute cloud velocity, $\alpha$ to be the angle between radial vector and velocity vector
and r$_{\rm cl}$ to be the cloud radius. L$_{\rm max}$ is the point with largest angular momentum 
and L$_{\rm min}$ the point with the smallest angular momentum (as long as the cloud is not engulfing the black
hole).
\label{motiv_a}}
\end{figure}

\begin{table}
\begin{center}
\begin{tabular}{cccc}
v$_{\rm cloud}$ [km/s] & $\frac{\rm N_{\rm engulf}}{\rm N_{\rm bypass}}$ & $\frac{\rm N_{\rm engulf}}{\rm N_{\rm total}}$ & $\frac{\rm N_{\rm bypass}}{\rm N_{\rm total}}$\\
\hline
20	&2.75	&2.17 \%	&0.79 \%	\\
30	&-	&0.75 \%	&0 \%		\\
50	&-	&0.006 \%	&0 \%		\\
80	&-	&0 \%	&0 \%		\\
\hline
\end{tabular}
\end{center}
\caption{Likelihood of the two scenarios of a cloud bypassing or engulfing the black hole during passage, calculated
for values suited for the Milky-Way GC. For a low cloud velocity of 20 km/s it is around three times more probable that 
a small disc will be formed by a cloud engulfing the black hole than by a cloud bypassing the black hole. For higher velocities
a bypassing cloud is no longer able to form a small disc at all. In the case of a high velocity cloud of 80 km/s neither of the
two scenarios is able to produce a small disc.
\label{motiv_table}}
\end{table}

In order to motivate our model we consider a cloud (modelled as a homogeneous sphere of gas)
on an arbitrary Keplerian orbit around the black hole.
We define d$_{\rm 0}$ to be the initial distance of the cloud from the black hole, v$_{\rm 0}$ to be the initial
absolute cloud velocity with respect to the stationary black hole, $\alpha$ to be the angle
between the radial vector and the velocity vector and r$_{\rm cl}$ to be the cloud radius (see Figure \ref{motiv_a}).
In addition we define the plane opened by the radial vector and the velocity vector to be the xy-plane in
Cartesian coordinates. We also restrict ourselves to $\alpha$ = 0 - $\frac{\pi}{2}$ due to symmetry.

The minimum/maximum angular momentum of the cloud is then given by the points on the cloud-surface in the xy-plane
for which the velocity vector is tangential to the cloud surface
(L$_{\rm z,min}$ will only be the minimum angular momentum as long as the cloud is bypassing the black hole).
The z-components of the specific angular momentum for these points are:

\begin{equation}
L_{\rm z,min} = +v_0 (r_{\rm cl} - sin(\alpha) d_0)
\end{equation}

\begin{equation}
L_{\rm z,max} = -v_0 (r_{\rm cl} + sin(\alpha) d_0) 
\end{equation}

All parts of the cloud will have the same direction of rotation if the sign of $L_{\rm z,min}$ is equal
to the sign of $L_{\rm z,max}$.
Otherwise, a part of the cloud will rotate in the opposite direction. Since all quantities are larger than
zero, the sign of $L_{\rm z,min}$ is given by the term $r_{\rm cl} - sin(\alpha) d_{\rm 0}$. If this term becomes
negative the whole cloud will bypass the black hole, if it becomes positive parts of the cloud will engulf the black hole
(and now L$_{\rm z,min}$ represents the maximum angular momentum of the counter-rotating part of the cloud).
Both of these scenarios can lead to the formation of an accretion disc.

In general, a clump of gas coming from infinity can still be captured by a black hole due to dissipation of 
its kinetic energy induced by the gravitational interaction with the black hole.
Shearing along the path of infall and tidal compression leads to colliding flows inside the clump
close to the black hole. This effectively turns kinetic energy
into thermal energy and thus reduces the cloud velocity. This way a previously unbound cloud can become bound to the black hole.
However, angular momentum needs to be conserved.
As long as the cloud stays on an eccentric orbit, tidal forces will lead to dissipation of kinetic energy during every
pericentre passage until the cloud finally ends up on a circular orbit with the same angular momentum as initially.
By comparing the angular momentum of a circular orbit to the initial angular momentum we can therefore 
calculate the radius of the final circular disc. Using $L_{\rm z,max}$ we get:

\begin{equation}
r_{\rm disk} = \frac{v_0^2 (r_{\rm cl} + sin(\alpha) d_{\rm 0})^2}{G M_{\rm BH}} 
\end{equation}

In order to estimate the likelihood of a cloud bypassing or engulfing the central black hole and thus forming the
observed stellar disc we restrict our parameters to values suited for the Milky-Way GC.
We take d$_{\rm 0}$ to be in the range of 5 - 50 pc and uniformly distributed. Here 5 pc corresponds to the centre 
of the circum-nuclear disc \citep{2003ANS...324..613V}, a reservoir of clumpy, molecular gas (around 10$^5$ M$_{\sun}$) which is
the first substantial source of gas near the black hole.
Following \citealt{2003ANS...324..613V} we restrict the source of our clouds to be within the central 50 pc region.
We already restricted $\alpha$ to 0 - $\frac{\pi}{2}$ and assume a Gaussian distribution
in $\alpha$ centred on $\frac{\pi}{4}$ since it
should be very unlikely that a cloud is on a direct radial orbit towards the black hole or that it originates
exactly from the apocentre of an elliptic orbit.
For the cloud velocity we take fixed values of 20, 30, 50 and 80 km/s.

We restrict the cloud radius r$_{\rm cl}$ to be in the range between 1.8 - 10 pc, which is motivated as follows.
Based on a GC region survey, \citealt{2000ApJ...536..357M} were able to derive the distribution of molecular cloud
sizes to be dN/dr$_{\rm cl}$ $\sim$ r$_{\rm cl}^{-4}$ which spans a range between 10 pc and their resolution limit of 3.3 pc.
Taking into account the total molecular gas of 10$^8$ M$_{\sun}$
in the inner 500 pc (\citealt{2004dimg.conf..253G}), the observed number of clouds of 160 and a
mean cloud mass of 10$^5$ M$_{\sun}$, we find that there should be roughly an additional
1000 clumps of 10$^5$ M$_{\sun}$.
We note that the required mass in the gaseous disc that formed the observed stellar disc
is around 10$^{5}$ M$_{\sun}$ \citep{2005A&A...437..437N, 2006MNRAS.366.1410N}.
Thus the smallest cloud radius, if we assume the distribution dN/dr$_{\rm cl}$ $\sim$ r$_{\rm cl}^{-4}$ to be
still valid, should be around 1.8 pc.
However, this assumption will strongly overestimate the number of small clouds
at the rather high mass of 10$^5$ M$_{\sun}$.
Our aim is to show that it is more likely that a cloud engulfs the black hole forming the observed stellar
disc than just bypassing it. Since small clouds are more likely to bypass the black hole we overestimate the amount
of bypassing clouds this way.
In addition, since e.g. a cloud of 10 pc radius cannot start at a distance d$_{\rm 0}$ smaller than 10 pc, we also
assume the upper limit on r$_{\rm cl}$ to be 3.5 pc at the smallest distance d$_{\rm 0,min}$ (d$_{\rm 0,min}$ - 3.5 pc 
gives roughly the inner edge of the circum-nuclear disc at 1.5 pc)
increasing linearly up to the maximum of 10 pc at the largest distance d$_{\rm 0,max}$.
We note that we exclude a number of possible formation scenarios such as the formation of a small disc by capturing
only part of a larger cloud or the possibility of a small stellar disc formed by a large accretion disc only fragmenting in the inner parts.
Thus, the discussion presented in this section only gives a rough estimate for the likelihood of such an event. A detailed study including 
hydrodynamics and also other formation models such as the plausible cloud-cloud collision model of \citealt{2009MNRAS.394..191H} is
beyond the scope of this paper.

We use a Monte-Carlo approach to sample values from the corresponding distributions, calculate the 
corresponding outer disc radius and check if the cloud would bypass or engulf the black hole. Tests have shown that
convergence in the first three digits after the decimal is reached after roughly 10$^8$ iterations.
We define N$_{\rm engulf}$ and N$_{\rm bypass}$ to be the total number of cases that lead to a small disc (r$_{\rm disk}$ $<$ 0.6 pc)
out of the total number of 10$^8$ values. The ratio N$_{\rm engulf}$/N$_{\rm bypass}$ for different cloud velocities 
determines the likelihood of the two scenarios. The ratio N$_{\rm engulf,bypass}$/N$_{\rm total}$ gives the absolute likelihood
of the individual scenarios.

Table \ref{motiv_table} summarises the results. For cloud velocities larger than 20 km/s the bypassing
scenario is not able to form a small disc at all. For the 20 km/s cloud it is still almost three times more
likely that the disc was formed by a cloud engulfing the black hole instead of a bypassing cloud.
At velocities of 30 km/s or higher only a cloud engulfing the black hole during passage is able to form a small disc.
The most prominent observed cloud near the GC has a velocity of 50 km/s. For higher cloud velocities such as 80 km/s
neither of the two scenarios can produce a small disc, which limits the velocity of the original cloud
that could have formed the observed disc to below 80 km/s.
The results can be explained by the fact that for the engulfing scenario we can have smaller impact parameters
and thus lower maximum angular momentum compared to the bypassing scenario. Thus even a rather large cloud
can lead to a small disc. With this we conclude that it is very likely that the cloud that formed the progenitor
gaseous accretion disc of the observed stellar disc, engulfed the black hole during its passage through the
very centre of the galaxy.


\section{Model and Numerical Method}
\label{num}

\subsection{Simulation Code}
\label{num_code}

The hydrodynamical evolution of the impact of the gas cloud with the central black hole
is studied using simulation runs performed with the N-body Smoothed Particle Hydrodynamics (SPH) Code Gadget2 
\citep{2005MNRAS.364.1105S}.
The code solves the Euler equations for hydrodynamics which neglect friction. To enable friction,
Gadget2 uses a modified version of the standard Monaghan, Gingold; Balsara
(\citealt{1983MNRAS.204..715G, 1995JCoPh.121..357B}) artificial viscosity:

\begin{equation}
\Pi_{ij} = -\frac{\alpha}{2} \frac{\left( c_{i} + c_{j} - 3 w_{ij} \right) w_{ij}} {\rho_{ij}} 
\end{equation}

with $w_{ij}={\vec{v}_{ij}\cdot\vec{r}_{ij}} / {\left|\vec{r}_{ij}\right|}$,
c$_i$ being the sound speed of particle i, 
v$_{ij}$, r$_{ij}$  the relative particle velocity and separation respectively, $\rho_{ij}$ the mean density and
$\alpha$ the artificial viscosity strength parameter.

The modification is necessitated, as the original form weights viscous forces strongly towards particles
with small separation. In the new formulation the induced pressure does not explicitly depend on the
hydro-smoothing length or particle separation. Tests have shown that in simulations with dissipation this has the
advantage of reducing very large viscous accelerations \citep{2005MNRAS.364.1105S}.
Gadget2 also includes an additional viscosity-limiter to remove spurious angular momentum transport.
This is implemented by using the mechanism proposed by \cite{1995JCoPh.121..357B} which adds a factor
(f$_i$ + f$_j$)/2 to the viscosity tensor, where

\begin{equation}
f_i= \frac{|\nabla \times \vec{v}|_i}{|\nabla \cdot \vec{v}|_i + |\nabla \times \vec{v}|_i }
\end{equation}
measures the shear around particle i.

It has been shown that the standard SPH
artificial viscosity produces a significant shear viscosity in rotationally-supported discs,
leading to spurious transport of angular momentum and that numerical dissipation is almost
never negligible and can be dominant
(\citealt{1995JCoPh.121..357B, 1996MNRAS.279..402M, 2004MNRAS.351..630L, 2006MNRAS.373.1039N}).
Thus we performed stability tests as shown in the Appendix to demonstrate that the above modifications are 
sufficient to make our results numerically stable and that we are not dominated by numerical viscosity.
For our standard simulations we used a value of $\alpha$ = 0.75. The tests shown in the Appendix \ref{numstab_alpha}
vary $\alpha$ in the range between 0.5 - 1.0 which is the
suggested range suited for numerical simulations \citep{2005MNRAS.364.1105S}.

\citealt{2005MNRAS.364.1105S} conducted a number of tests to show that Gadget2 is able to resolve strong
shocks.
The main difference of SPH codes compared to grid based codes is that droplets of gas which form
during shocks can survive 
rather long and do not mix as efficiently with the background material as in grid codes (see e.g.
\citealt{1993A&A...268..391S, 2007MNRAS.380..963A}). This is due to the fact that in SPH the forces
at the edges between two phases are smoothed out since there are always contributions from both SPH particles
inside the clump and from SPH particles in the surrounding medium. 

In order to carefully account for the thermodynamics of disc formation, we implemented the cooling formalism of 
\cite{2007A&A...475...37S} into Gadget2.
It approximates cooling processes for optically thin, as well as optically thick regions by applying a
approximative radiative transfer mechanism using the diffusion approximation.
The basic equation determining the cooling rate is given by:

\begin{equation}\label{radcool}
\frac{du}{dt} = \frac{4\,\sigma_{_{\rm SB}}\,(T_{_{\rm 0}}^4 -T^4)}
{\bar{{\Sigma}}^2\,\bar{\kappa}_{_{\rm R}}(\rho,T)\,+\,\kappa_{_{\rm P}}^{-1}(\rho,T)}
\end{equation}
with $T_0$ the background radiation temperature, T and $\rho$ the particle temperature and density,
$\bar{{\Sigma}}$ the pseudo-mean surface-density, $\bar{\kappa}_{_{\rm R}}$ the pseudo-mean opacity and
$\kappa_{_{\rm P}}$ the Planck-mean opacity. We refer to \cite{2007A&A...475...37S} for the definition and interpretation
of the pseudo-mean surface-density and the pseudo-mean opacity, which are crucial for their method in approximating
radiative transfer (see Equations 18 and 23 in \citealp{2007A&A...475...37S}).

The method uses the particle density, temperature and gravitational potential to estimate a mean optical depth for each SPH particle which
then regulates its heating and cooling.
For a small optical depth, Eqn. \ref{radcool} reduces to the usual cooling term for the optically thin cooling regime.
\begin{eqnarray}
\frac{du}{dt} \simeq -4 \sigma_{_{\rm SB}} T^4 \kappa_{_{\rm P}}(\rho,T)
\end{eqnarray}
For a large optical depth, Eqn. \ref{radcool} reduces to the diffusion approximation (see Eqn. 27 and 28 in \cite{2007A&A...475...37S}).

At each timestep the following steps are undertaken:
\begin{itemize}
\item Approximate the mean optical depth of the environment of each SPH particle by using the density and gravitational potential
      following the procedure as described in \cite{2007A&A...475...37S}.
      In the case of non-SPH particles, exclude them from the gravitational potential calculation.
\item Calculate the compressive heating rate and radiative cooling rate. In the case of T $<$ T$_0$ we have radiative heating 
      instead of cooling.
\item Update the internal energy and temperature of each SPH particle according to the total change in heating plus cooling.
\end{itemize}

In this cooling formalism the molecular weight is temperature dependent, taking into account the degree of dissociation and ionisation of hydrogen
as well as the degree of single and double ionisation of helium.
The specific internal energy of an SPH particle is the sum of contributions from molecular, atomic and ionised hydrogen
as well as the contributions from atomic, single and double ionised helium together with the associated dissociation energies.
The method is suited for a wide range of temperatures (T = 10 - 10$^6$ K), a
wide range of densities ($\rho$ = 10$^{-19}$ g/cm$^3$ - 10$^{-2}$ g/cm$^3$) and optical
depths in the range of 0 $< \tau <$ 10$^{11}$ as shown in tests performed by \cite{2007A&A...475...37S}.  
The additional computational overhead for this method is not too high, since it can primarily be implemented through
look-up tables and requires only a few real-time calculations.


\subsection{Numerical Parameters}
\label{num_para}

The simulations are run with a total number of $10^6$ SPH particles.
To test the resolution dependence of our results we did one very high resolution simulation
with 5$\times$10$^6$ SPH particles presented in the Appendix \ref{numstab_res}.
The number of SPH neighbours is set to $n_{\rm neigh} = 50\pm 5$.
All simulations use a gravitational softening length of $\epsilon = 10^{-3}$ pc.
In what follows we only discuss the gravitational softening length since the code only allows to change
this value directly and sets the minimum hydro-smoothing length to a fraction of this value. In our case we choose the minimum hydro-smoothing
length to be equal to the gravitational softening length according to \citealt{1997MNRAS.288.1060B}.
We use the standard Gadget2 implementation of a fixed gravitational softening length and a
variable hydro-smoothing length, thus the gravitational softening length is always $\epsilon$ = 10$^{-3}$ pc.
The softening length strongly influences the simulation runtime, with a lower value leading to a longer runtime.
To test the dependence of our results on softening length we performed test simulations
with a softening length of $\epsilon = 10^{-4}$ pc. The outcome of this test is presented in the Appendix \ref{numstab_soft}.

In order for the simulations not to become too time-consuming as a result of very
small particle time steps in the vicinity of the black hole we define an
accretion radius $r_{\rm acc}$ within which all SPH particles are considered to be accreted
by the black hole. These particles are removed from the simulation and will no longer take part
in the dynamical evolution.
The fixed-position sink is not associated with a particle.
This mechanism does not strictly conserve linear or angular momentum,
but due to the small amount of accreted gas (at most 1\% of the black hole mass)
the violation of angular momentum conservation is negligible.

We want the accretion radius $r_{\rm acc}$ to be larger than the softening
length of $10^{-3}$ pc and smaller than the observed inner disc edge of $4\times 10^{-2}$ pc.  
Test simulations showed that a value of $r_{\rm acc}=2\times 10^{-2}$ pc allows
us to simulate the encounter within a reasonable total runtime such that the whole cloud has passed the black hole and an compact accretion disc
has formed.

We investigate three different types of simulations. The first type of simulations are purely adiabatic simulations
with cooling only due to adiabatic expansion. The second type of simulations are isothermal simulations and 
finally, representing the most realistic
case, we performed simulations using the full cooling prescription of \cite{2007A&A...475...37S} presented
in section \ref{num_code}.

\subsubsection{Adiabatic simulations}
\label{num_para_ad}
Using the ideal gas approximation we have an equation of state of $P = \frac{\rho k_B T}{\mu m_H}$ with
$\rho$, T gas density and temperature respectively, $k_B$ the Boltzmann-constant, $m_H$ the hydrogen mass
and the molecular weight $\mu = 2.35$ assuming a mixture of 70\% hydrogen and 30\% helium (molecular).
The initial temperature is set to T$_{\rm cloud}$ = 50 K which is a typical temperature for a cloud in the GC
\citep{2004dimg.conf..253G}. 

\subsubsection{Isothermal simulations}
\label{num_para_iso}
The isothermal simulations represent the case for which cooling is balanced by heating. Given our assumed low
temperatures, it corresponds to very efficient cooling. 
The isothermal simulations use an equation of state of $P = c_s^2 \rho$ with $\rho$ the gas density and
$c_s$ the sound speed of the gas, calculated initially by using $c_s^2 = \frac{k_B T}{\mu m_H}$.
Again the initial temperature is set to 50 K.

Heating and cooling processes in the disc will however play an important role when
investigating the fragmentation and condensation of the disc into stars (see e.g. \citealt{2008Sci...321.1060B}).
Observations have shown that the stellar disc IMF is very top-heavy \citep{2010ApJ...708..834B},
indicating that a large Jeans-mass is required for the forming stars, which in term implies that cooling should not
be too efficient. On the other hand the temperatures have to remain low enough for the 
system to form. If compressional heating becomes too efficient the strong expansion of the gas
prevents the formation of the accretion disc, as our purely adiabatic simulations show.

\subsubsection{Full Cooling simulations}
\label{num_para_cool}
Our third set of simulations uses the full cooling
prescription of \cite{2007A&A...475...37S} presented in section \ref{num_code}.
The ideal gas approximation is used again, assuming initial cloud temperature $T_{\rm cloud}$ and background radiation
temperature $T_0$ (due to the old spherical cluster of stars near the GC) to be 50 K. The
molecular weight is a function of temperature and takes into account the processes described in section \ref{num_code}.


\subsection{Initial Conditions}
\label{num_initial}

\begin{table}
\begin{center}
\begin{tabular}{cccccc}
		ID & \imp & \velp & $v_{\rm 0}$ [km/s] & b [pc] & j$_{\rm specific}$ [$\frac{\rm pc^2}{\rm Myr}$]\\
\hline
        I01 & 0.28 & 0.29 & 50  & 1 & 50\\
        I02 & 0.57 & 0.41 & 50  & 2 & 100\\
        V01 & 0.85 & 0.3  & 30  & 3 & 90\\
        V02 & 0.85 & 0.5  & 50  & 3 & 150\\
        V03 & 0.85 & 0.8  & 80  & 3 & 240\\
	V04 & 0.85 & 1.0  & 100  & 3 & 300\\
	C01 & 0.57 & 0.98 & 120 & 2 & 240\\        
\hline
\end{tabular}
\end{center}
\caption{Summary of initial conditions for our simulations. The cloud velocity is varied
for V01, V02, V03 and V04. The impact parameter is varied for I01, I02 and V02.
C01 is a comparison setup with the same initial specific angular momentum (j$_{\rm specific}$) as V03.
\label{num_itable}}
\end{table}

In this first study we start with a spherical and homogeneous cloud of radius 3.5 pc which is typical for the
GC (\citealt{2000ApJ...536..357M}, derived from CS(1-0) radio surveys). The $H_{2}$ gas density
is $10^4$ $\mathrm{cm^{-3}}$ (\citealt{2004dimg.conf..253G}, taken from high-resolution surveys of CS and
CO in the GC), leading to a cloud mass of $8.81\times 10^{4}$ M$_{\sun}$. The SPH particle
mass is then $m_{\rm SPH} = 8.81\times 10^{-2}$ M$_{\sun}$ and the corresponding minimum mass that can be
resolved is $m_{\rm min} = n_{\rm neigh} \times m_{\rm SPH} = 4.4$ M$_{\sun}$ (following \citealp{1997MNRAS.288.1060B}).

The black hole is included as a static potential of a point mass of $M_{\rm BH}= 3.5\times 10^6$ M$_{\sun}$
\citep{2003ApJ...594..812G} placed at the origin of the coordinate system.
We do not include black hole growth in order to speed up the simulations. 
This growth is anyhow negligible since at most only around one percent of the black hole mass was accreted
at the end.
Energetic feedback from the accreting black hole, (see e.g. \citealp{Johansson:2009BH}) might however be important
and will be investigated in a subsequent paper.
The Gadget2 Code treats gravitational forces as Newtonian up to 2.8 times the
gravitational softening length, which in any case is beyond our accretion radius for the softening length.

Initially, the centre of mass of the cloud is placed at a distance d$_{x0}$=5 pc from the
origin/black hole on the x-axis (the direction of motion) and an offset $b$ on the y-axis.
Note that compared to the coordinate system used in the motivation (Sec. \ref{motiv}) where
we defined the x-axis to be along the line connecting the centre of the cloud with the black hole, we now
define the x-axis to be parallel to the cloud velocity vector passing through the coordinate origin/black hole position.
This way the cloud to black hole distance d$_0$ defined in the motivation section relates to d$_{x0}$ and $b$ by
$d_0^2 = b^2 + d_{x0}^2$.

We have chosen this relatively small distance in order to prevent the cloud from collapsing before
it reaches the black hole. Additionally this is also the distance of the so called circum-nuclear
disc (a reservoir of clumpy, molecular gas of around 10$^5$ M$_{\sun}$) from the GC
\citep{2003ANS...324..613V} for which there is evidence for infall of gas towards the GC
\citep{2009ApJ...695.1477M}.
The origin of gas with low enough angular momentum to feed the GC in general
(e.g. \citealt{2007MNRAS.377L..25K}) is still an area of active research and
beyond the scope of this paper.
 
The free fall time of a cloud with $n=10^4$ $\mathrm{cm^{-3}}$ is only around $0.36$ Myrs.
For an average infall velocity of 50 km/s\, the cloud's centre of mass should
reach the black hole after roughly 0.1 Myrs (assuming no acceleration due to the black hole).
The evolution is characterised by the fraction \imp\, where
r$_{\rm cl}$ is the cloud radius and $b$ is the initial cloud's centre of mass offset on the y-axis.
\imp\ is always smaller than one in our case of a cloud engulfing the black hole.

The initial cloud velocity v$_0$ in (negative) x-direction is a typical value, taken from
the observations presented by \citet{2000ApJ...536..357M}.
The evolution is then characterised by the ratio \velp\, where v$_0$ is the initial cloud velocity and 
$v_{b}^{2} = \frac{2G M_{bh}}{b}$ is the black hole's escape velocity at distance $b$.
The initial conditions of our parameter studies are shown in Table \ref{num_itable}.

Three setups (I01, I02, V02) use a fixed initial cloud velocity of 50 km/s
and varying offsets $b$, set to 1, 2 and 3 pc, respectively. In addition
there are four setups (V01, V02, V03, V04)  with $b$ fixed to 3 pc and varying initial cloud velocities,
set to 30, 50, 80 and 100 km/s, respectively.
The comparison setup C01 has the same initial specific angular momentum as V03
but a smaller impact parameter and higher velocity.

We do not include the potential of the massive stellar cusp in the GC \citep{2003ApJ...594..812G} into our simulations
which would lead to precession of the orbits on a time-scale that is short compared to the infall time
\citep{2005MNRAS.358.1361I}.
In this first step we only concentrate on the dynamics due to the black hole cloud interaction and postpone 
investigation of the effects of including the stellar cusp to the next paper in our series. 

The simulations were typically run for an evolutionary time of 0.25 Myrs which corresponds 
roughly to the instant when all parts of the cloud have crossed the black hole.
Taking into account that disc formation is finished after roughly 0.15 Myrs, when the disc becomes almost static,
we evolve the disc for a bit less than 2 full orbits at a radius of 1 pc. During this short time period, viscous evolution does not
play any significant role. 

Only simulation I01 was stopped already after 0.2 Myrs. At that time, due to the small
impact parameter, the time step became very small making the simulation very time-consuming and too
expensive to continue. Again we want to stress that in this paper we are only interested in the resulting gaseous
disc properties directly after the collision with the black hole without following the viscous evolution of the
accretion disc and we defer the study of fragmentation and star formation to a follow-up paper.


\section{Results}
\label{results}

\subsection{The Adiabatic Case}
\label{ad_res}

\begin{figure*}
\begin{center}
\includegraphics[width=18cm]{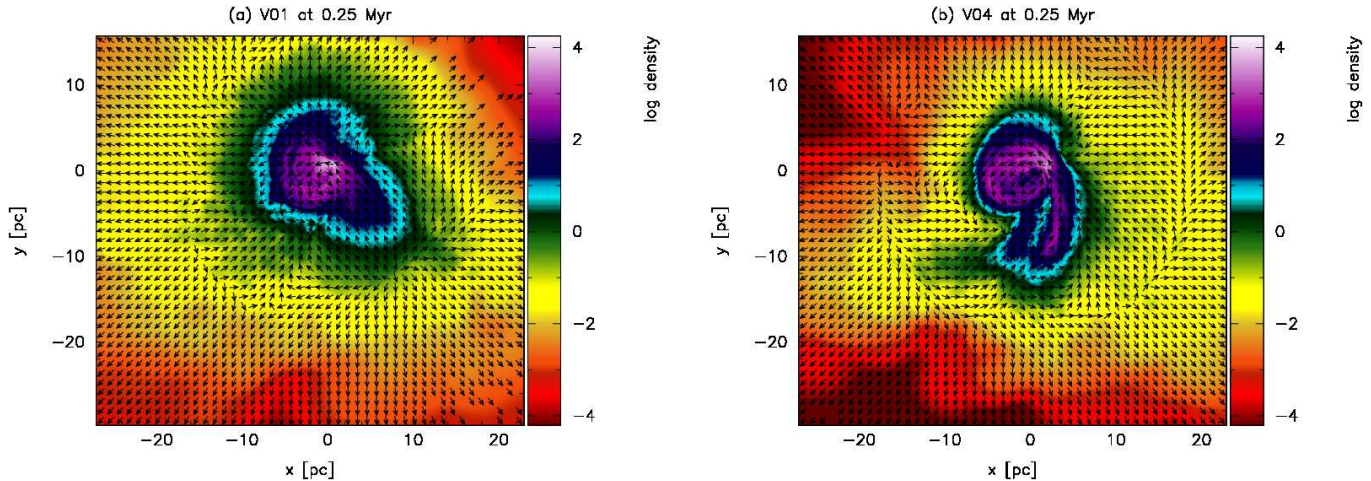}
\end{center}
\caption{Logarithmic density in units of $\frac{M_{\sun}}{\mathrm{pc^3}}$ in the xy-plane with z=0 pc
for purely adiabatic simulations using initial conditions V01 (a) and V04 (b) shown at the final time step
of t=0.25 Myrs. Velocity vectors are all set to the same length and only indicate the local direction.
Clearly visible in (a) and (b) is the spherically expanding shell of gas due to the strong adiabatic heating. 
In addition, an extended ring of gas formed in the inner 5 pc around the black hole.
\label{ad_evol}}
\end{figure*}

\begin{figure}
\begin{center}
\includegraphics[width=9cm]{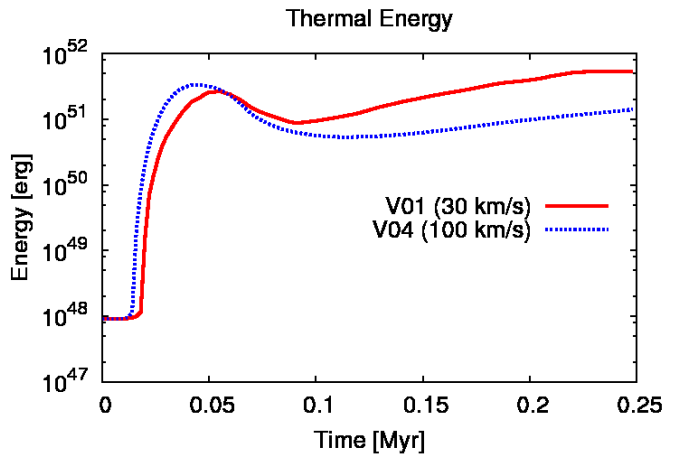}
\end{center}
\caption{Evolution of total thermal energy as a function of time for the two adiabatic simulations, which increases
nearly four orders of magnitude.
This is already comparable with supernova energies of typically around 10$^{51}$ erg.
Shown are the cases for initial condition V01 (red, solid line) and V04 (blue, dotted line).
\label{ad_energy}}
\end{figure}

The first simulations we present are purely adiabatic without cooling and use initial conditions V01 and V04,
which differ only in initial cloud velocity with 30 km/s for V01 and 100 km/s for V04.
The resulting densities in units of $\frac{M_{\sun}}{\mathrm{pc^3}}$ in the xy-plane for z=0 pc at the end of the
simulations (t=0.25 Myr) are shown in Figure \ref{ad_evol}.
All velocity vectors are set to the same length and thus only represent the corresponding local direction
of the velocity field.

The gas is strongly spherically expanding as a result of the heating due to adiabatic compression
during the initial impact.
This mechanism basically completely prevents the formation of a disc.
Only an extended ring-like structure evolves inside the inner 5 pc as seen in Figure \ref{ad_evol}.
A huge bubble of hot gas with a radius of roughly 30 pc forms around the ring
which is still expanding at the end of the simulation.

Figure \ref{ad_energy} shows the increase in thermal energy over time for both simulations.
The thermal energy for the 100 km/s cloud rises by a larger amount compared to the 30 km/s
cloud as expected due to the stronger shocks. At a later point the expanding gas starts to cool because of
adiabatic expansion. Since there is more gas initially
bound to the black hole in the case of the 30 km/s cloud, more material is still in the centre and thus
shocks inside the surviving ring generate more heat compared to the faster cloud.
The final energy can rise up
to several times 10$^{51}$ erg and becomes comparable with supernova energies of typically 10$^{51}$ erg. 
From this we
conclude that if cooling becomes too inefficient, the interaction of a cloud with the central super-massive black hole
will lead to an explosion and the formation of a hot, expanding bubble with energies comparable to supernova
explosions. However this is a very unlikely
scenario since high gas-densities typically lead to strong cooling.


\subsection{The Isothermal and Full Cooling Cases}
\label{icool_res}

\begin{figure*}
\begin{center}
\includegraphics[width=18cm]{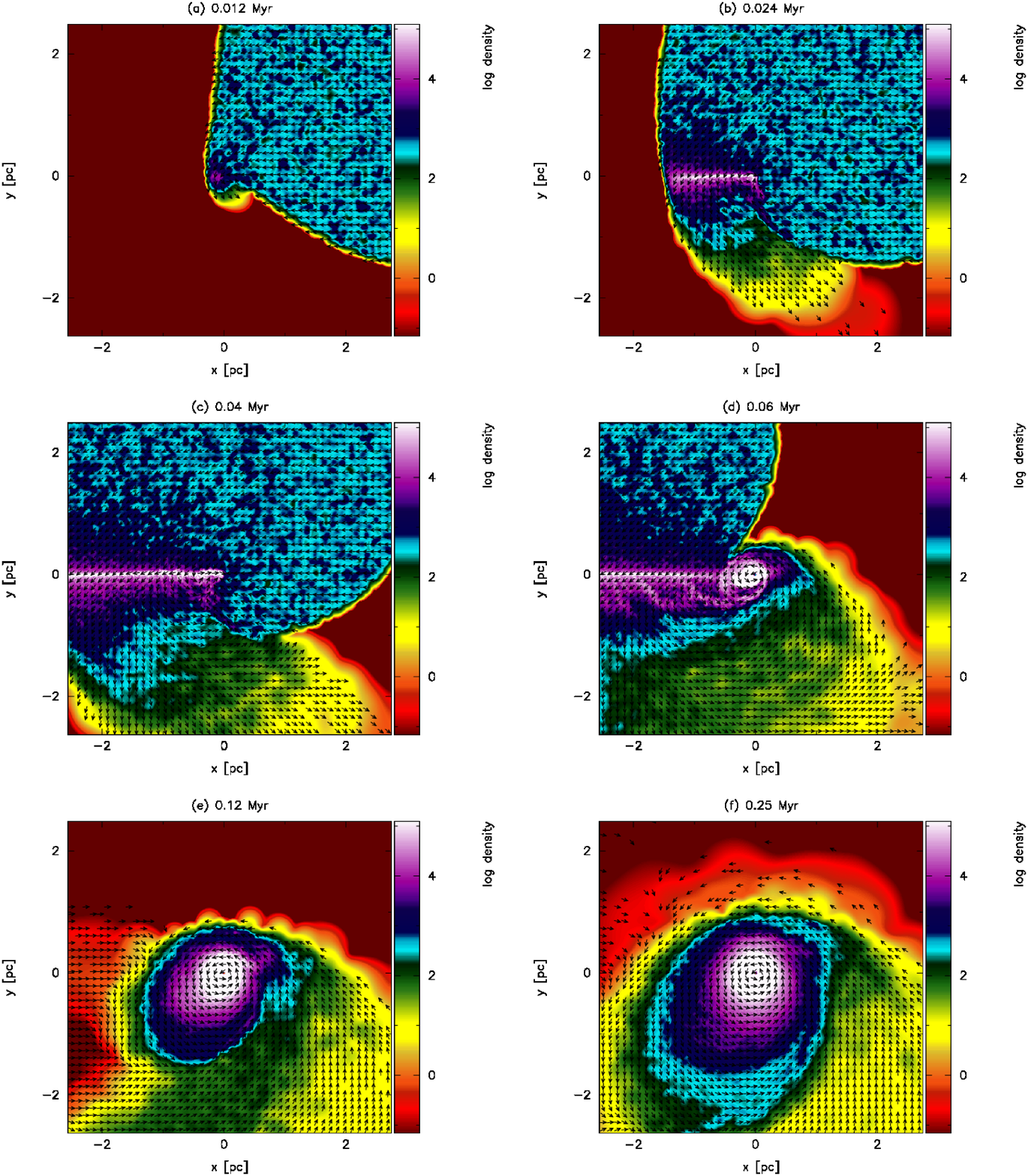}
\end{center}
\caption{Logarithmic density in units of $\frac{M_{\sun}}{\mathrm{pc^3}}$ in the xy-plane with z=0 pc at
different times using initial condition C01 with cooling enabled. Velocity vectors are all the same length
and only indicate local direction. Gas rotates around the black hole from opposite directions
and collides, leading to a high density region along the line of impact between the two streams.
The high density region seen in b,c,d along the negative x-axis
is created by material with a large initial distance in z-direction, concentrated into the z=0 pc plane.
Global rotation is counter-clockwise due to the initially larger amount of gas on counter-clockwise
orbits, however due to the interaction at the high density region more gas becomes bound to the
black hole than expected from the initial setup. C01 represents an extreme case in which there
should be nearly no gas bound to the black hole at all. Thus, the formation of an accretion disc supports
our proposed mechanism.
\label{icool_evol}}
\end{figure*}

\begin{figure}
\begin{center}
\includegraphics[width=9cm]{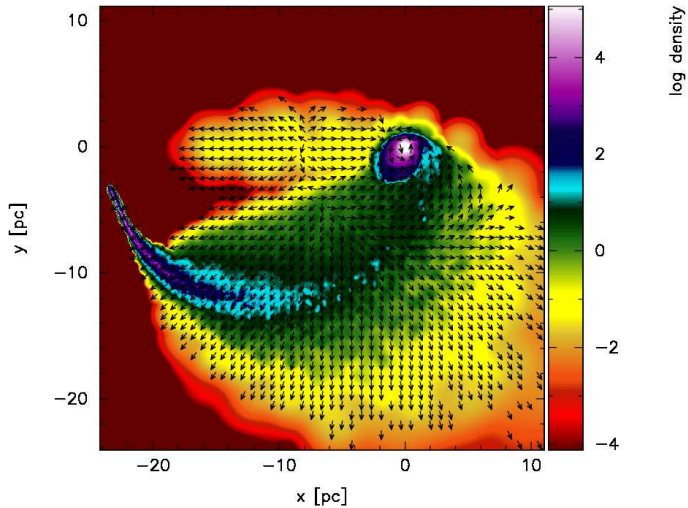}
\caption{A zoom-out of the logarithmic density plot from Figure \ref{icool_evol}f
in units of $\frac{M_{\sun}}{\mathrm{pc^3}}$ in the xy-plane with z=0 pc. Velocity vectors again only indicate
the local direction, all vectors are set to the same length.
Clearly visible is the escaping part of the cloud in the lower-left part of the plot, as well as material
still falling onto the accretion disc.
\label{icool_evollarge}}
\end{center}
\end{figure}

\begin{figure*}
\begin{center}
\includegraphics[width=18cm]{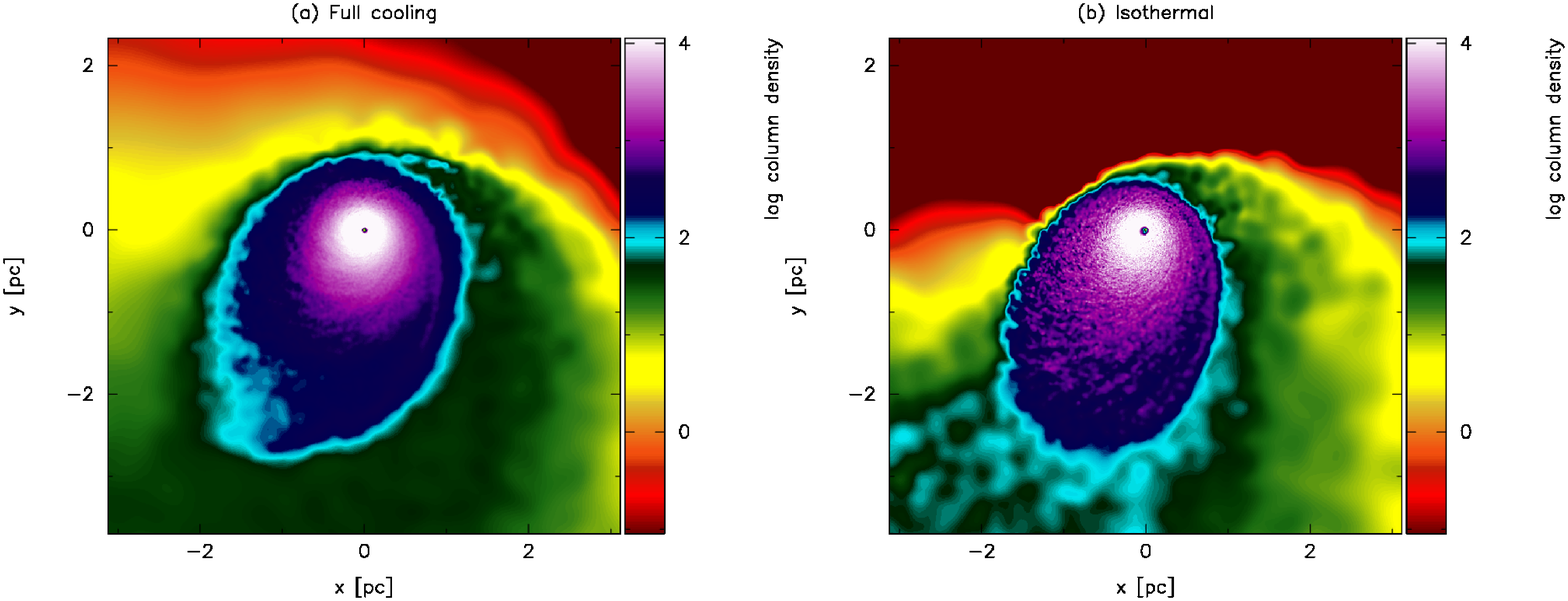}
\caption{A comparison of the surface-density
in units of $\frac{M_{\sun}}{\mathrm{pc^2}}$ in the xy-plane for the full cooling case (a) and the
isothermal case (b) at the final timestep of 0.25 Myrs using initial condition C01.
Both discs are very similar in terms of size and shape. As expected the isothermal disc shows stronger
signs of condensation compared to the hotter disc from the full cooling simulation.
\label{icool_comp}}
\end{center}
\end{figure*}

We used initial conditions I02, V01, V02, V03 and C01 for the isothermal and the full cooling cases.
As already mentioned, I01 is a very time-consuming setup and we have only run this setup for the 
computationally faster isothermal case.

\subsubsection{The Isothermal Case}
From a simulation standpoint
an isothermal equation of state is easy to implement and thus a good starting point in a number of cases.
Judging from our results in this case compared to the more realistic case using the full cooling prescription
we can say that at least some results can already be obtained with the 
isothermal approximation. Still one has to keep in mind that there are problems with the isothermal approach, which
we will show in the following sections and which we already mentioned in section \ref{num_para_iso}.

First we note that with an isothermal equation of state fragmentation is artificially enhanced because of our extremely
low temperatures. However, we do not yet include star formation and are only interested in the properties of the
forming gaseous disc.
Another point is that in all isothermal simulations, the SPH particles in the inner 0.1 pc reach the minimum
softening length so that we are limited by numerics in this region and the physics is not resolved 
due to the strong softening over the forces and thus we cannot make any reliable 
predictions about the disc properties here.

The time evolution of surface density of the discs in the xy-plane in the isothermal case is very similar to the
full cooling case
(as shown for C01 in Fig. \ref{icool_comp})
we will present next. However the isothermal discs are very
flat in the z-direction due to the low temperature
in comparison to the full cooling discs. Because of this the evolution of density in the xy-plane for z=0 pc differs
for the isothermal and full cooling discs, with the full cooling discs obviously showing lower densities. 
Still the evolution and general shape of the forming disc
as presented for the full cooling case in the next section can also be taken as representative for the isothermal discs.

\subsubsection{The Full Cooling Case}
The time evolution of density in units of $\frac{M_{\sun}}{\mathrm{pc^3}}$ in the xy-plane (z=0 pc)
for initial condition C01 can be seen in Figure \ref{icool_evol} for the cooling case.
Velocity vectors are all set to the same length and only represent the local direction in the xy-plane.
We have chosen to show C01 (cloud velocity of 120 km/s and impact parameter of 2 pc) since it represents an
extreme case in which nearly no gas is bound to 
the black hole initially due to the high cloud velocity. Nonetheless, we form a gaseous accretion disc 
around the black hole at the end of the simulation.

After starting the simulation, the cloud approaches the black hole from the right hand side along the x-axis.
Due to tidal forces the parts closest to the black hole quickly start to form a finger-like extension
stretching towards the black hole located at the coordinate origin as seen in Figure \ref{icool_evol}a. 

The gas passes the black hole on orbits corresponding to its initial angular momentum.
12000 yrs later (Figure \ref{icool_evol}b), this material begins to collide with gas streaming around the black hole
from the opposite side.
This leads to a high density region along the line of impact between the two streams.
The high density region along the negative x-axis, as seen in Figures \ref{icool_evol}b,c and d,
is created by material with a large initial distance in z-direction, concentrated into the z=0 pc plane.

Due to the larger amount of gas initially rotating counter-clockwise the overall rotation follows this direction, 
but the interaction with the clockwise rotating gas removes kinetic energy
and lowers the angular momentum, forcing the gas on bound orbits
or to fall directly into the black hole.
This bound gas builds up the accretion disc as seen in Figures \ref{icool_evol}d,e.

The simulation was stopped after 0.25 Myrs when the cloud passed around the black hole. An inner gaseous accretion disc
has formed at that time.
This evolutionary state is shown in Figure \ref{icool_evol}f.
Figure \ref{icool_evollarge} shows a larger region around the final state shown in Figure \ref{icool_evol}f.
Clearly visible is the escaping part of the cloud below -10 pc on the x-axis and at around
-10 pc on the y-axis, as well as material still falling onto the accretion disc.

In general other initial conditions show similar behaviour as described above.
They differ in the amount of gas which is already
bound to the black hole initially due to lower cloud velocities and thus lead to more massive discs.
For larger impact parameters the amount of gas on clockwise orbits is smaller and thus the interaction inside
the region where the counter-rotating streams collide is not as strong as in the cases with smaller impact parameter. This
can also be seen when we discuss accretion later in section \ref{icool_accretion}. 
Cloud velocity determines the final rotation of the resulting discs semi-major axis with respect to the x-axis
in the xy-plane. For a 50 km/s cloud the semi-major axis aligns directly with the x-axis, whereas for a
80 km/s cloud it is rotated towards the negative y-axis and for a 30 km/s cloud towards the positive y-axis.


\subsection{Eccentricity and Mass Distribution}
\label{icool_ecc}

\begin{figure*}
\begin{center}
\includegraphics[width=18cm]{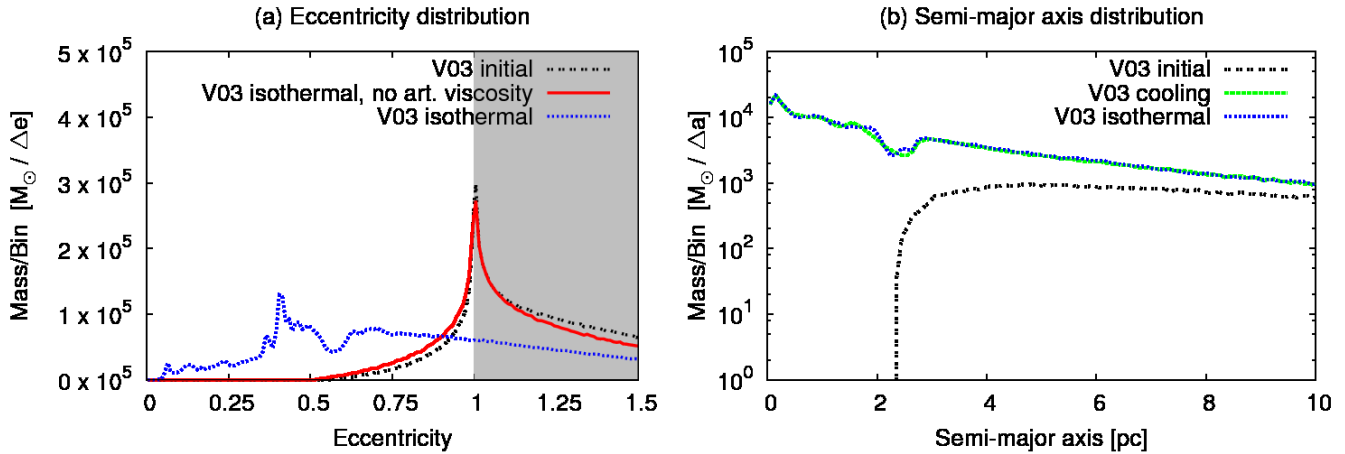}
\end{center}
\caption{Comparison of the distribution of mass onto orbits of different eccentricity (a) and semi-major axis
distribution (b). The black (dash-dotted) line in (a)
shows the expected distribution from initial condition V03. Particles with e $>$ 1 are not bound to the black hole,
indicated by the grey background.
The red (solid) line shows an isothermal simulation without artificial viscosity using initial condition V03
at the final time step of t=0.25 Myrs. In this run there are no large pressure gradients 
during the simulation so that the change in mechanical energy and 
eccentricity is small. The final distribution thus approximately reflects the initial conditions.
The blue (dotted) curve shows the same simulation (isothermal and using initial condition V03)
at t=0.25 Myrs with artificial viscosity now turned on. In this case significantly more mass settles into closed, more
circular orbits than expected from the initial conditions. 
In (b) the black (dash-dotted) line represent the distribution of mass onto orbits with different semi-major axis for initial
condition V03, compared to the isothermal (blue, dotted) and full cooling case (green, dashed) at the end of the simulations.
A significant amount of gas settles into orbits very close to the black hole. Table \ref{icool_angular_tab} summarises
the cancellation of angular momentum which leads to the formation of the compact (semi-major axis $<$ 1pc) disc.
\label{icool_ecc_a}}
\end{figure*}

\begin{figure*}
\begin{center}
\includegraphics[width=18cm]{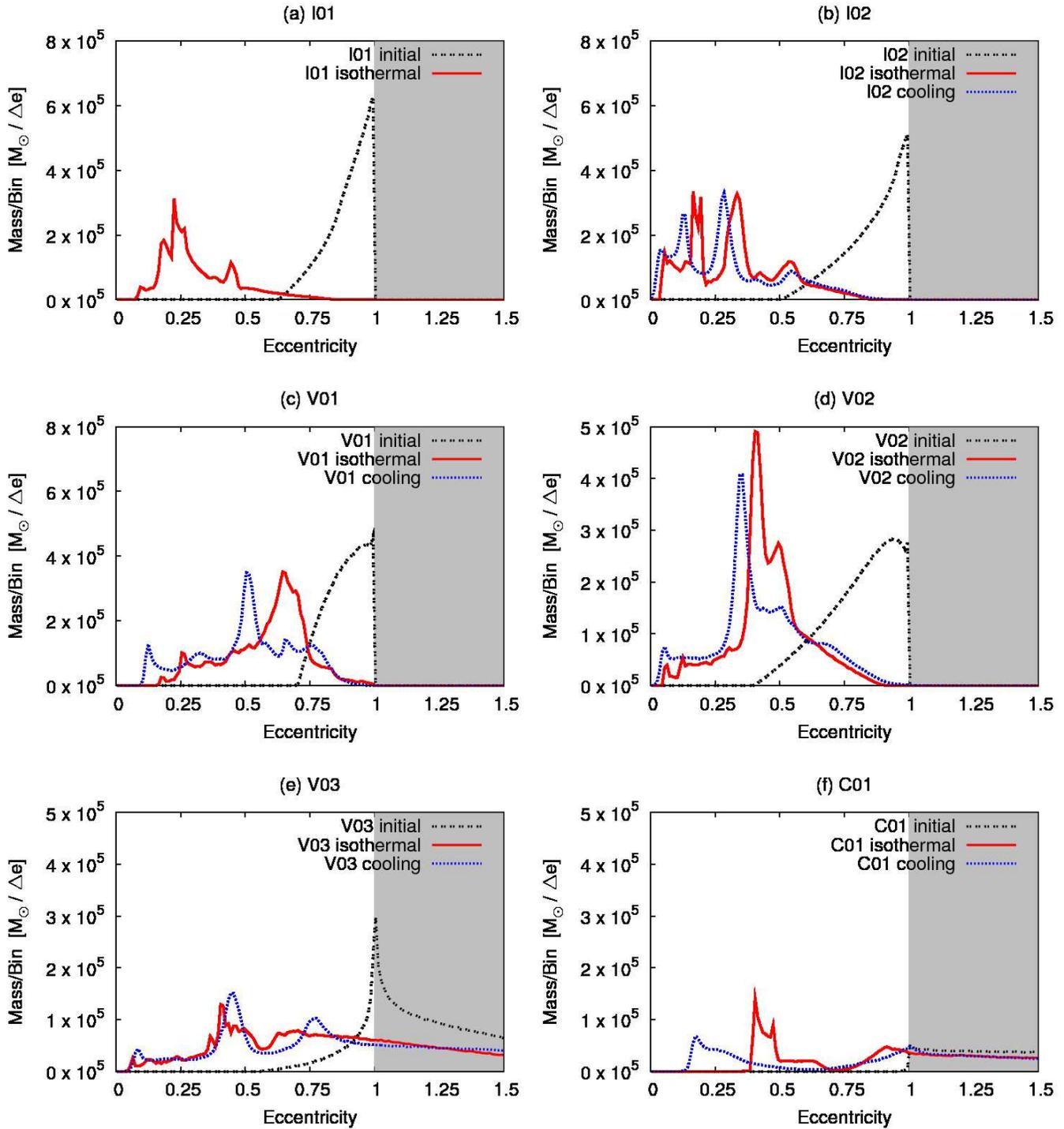}
\end{center}
\caption{Comparison of the distribution of mass onto orbits of different eccentricity for all initial conditions
using isothermal simulations and adiabatic simulations with cooling.
Particles with e $>$ 1 are not bound to the black hole, indicated by the gray background.
The black line always shows the distribution 
resulting from the initial conditions. The red (solid) line shows the distribution for the isothermal simulations and the
blue (dotted) curve the distribution for the simulations with cooling. In general, the more realistic full
cooling simulations tend to form more circular discs compared to the isothermal case. Still the isothermal simulations
already yield results similar to the cooling case for most initial conditions.
\label{icool_ecc_b}}
\end{figure*}

\begin{table}
\begin{center}
\begin{tabular}{cccc}
		ID & $M_{\rm acc, exp}$\footnotemark[1]  & $M_{\rm acc, iso}$\footnotemark[1] & $M_{\rm acc, cool}$\footnotemark[1]\\
\hline
	   I01 & 2.45  & 40.07  &   -   \\
	   I02 & 2.10  & 11.90  & 13.03 \\
           V01 & 3.14  &  4.72  &  5.27 \\
	   V02 & 1.26  &  2.51  &  2.71 \\
	   V03 & 0.50  &  1.57  &  1.76 \\
	   C01 & 0.37  & 11.32  & 11.85 \\
\hline
\end{tabular}
\end{center}
\footnotemark[1]All masses in units of 10$^3$ M$_{\sun}$
\caption{The accumulated mass $M_{\rm acc, exp}$ on orbits within the accretion radius as expected from the initial conditions 
compared with the result $M_{\rm acc,iso}$ for the isothermal simulations and $M_{\rm acc,cool}$ for the full cooling
simulations after the final time step
of t=0.25 Myrs. The isothermal simulation using initial condition V03 without artificial viscosity 
presented in Figure \ref{icool_ecc_a} accreted a mass of 0.49 $\times$ 10$^3$ M$_{\sun}$ at the final time step of
t=0.25 Myrs compared to 0.5 $\times$ 10$^3$ M$_{\sun}$ expected from initial conditions. This clearly demonstrates that
the large difference in expected accreted mass to actual accreted mass in the table is due to the interaction of
the two streams of gas around the black hole which does not happen in the special simulation without artificial viscosity.
\label{icool_ecc_tab}}
\end{table}

\begin{table}
\begin{center}
\begin{tabular}{cccc}
ID & $M_{\rm bnd, ini}$\footnotemark[1] & $M_{\rm bnd, iso}$\footnotemark[1] & $M_{\rm bnd, cool}$\footnotemark[1]\\
\hline
	   I01 & 88.10 & 48.03  & -\\
	   I02 & 88.10 & 76.20  & 85.20 \\
           V01 & 88.10 & 83.37  & 82.82 \\
	   V02 & 88.10 & 85.58  & 85.35 \\
	   V03 & 15.50 & 51.53  & 49.78 \\
	   C01 & 0.84  & 19.03  & 17.61 \\
\hline
\end{tabular}
\end{center}
\footnotemark[1]All masses in units of 10$^3$ M$_{\sun}$
\caption{The total initially bound (eccentricity e $<$ 1) mass $M_{\rm bnd, ini}$ compared to the total bound mass at the end
of the simulations (0.25 Myrs) for the isothermal case $M_{\rm bnd, iso}$ and the full cooling
case $M_{\rm bnd, cool}$. Only for initial conditions C01 and V03 not all the mass (88.1 $\times$ 10$^3$ M$_{\sun}$)
is already bound to the black hole initially. Due to accretion the bound mass is lower than the initial value for 
the initial conditions that have all mass bound to the black hole initially. This represents just the difference between initial total mass
and final total accreted mass. 
For initial conditions C01 and V03 there is a significant fraction of the initial total mass that got bound
to the black hole (20\% - 40\%) and a part that is still unbound and escaping.
\label{icool_bound_tab}}
\end{table}

\begin{table}
\begin{center}
\begin{tabular}{cccc}
ID & Isothermal\footnotemark[1] & Full cooling\footnotemark[1] & Compact disc\footnotemark[1] \\
\hline
	   I01 & 0.58 & -    & 0.47 \\
	   I02 & 0.76 & 0.76 & 0.67 \\
           V01 & 0.89 & 0.90 & 0.87 \\
	   V02 & 0.87 & 0.88 & 0.81 \\
	   V03 & 0.85 & 0.86 & 0.70 \\
	   C01 & 0.72 & 0.75 & 0.37 \\
\hline
\end{tabular}
\end{center}
\footnotemark[1]Mean of the ratio of angular momentum $\frac{|J_{\rm final}|}{|J_{\rm initial}|}$
\caption{
Ratio of the absolute value of angular momentum at the final timestep to the
absolute value of initial angular momentum, averaged over all particles for the isothermal and the full cooling case.
Only particles which are present in the initial and the final snapshot are compared. This neglects already accreted particles 
(which make up 2-15\% of the total initial particle number)
since they are no longer present in the final snapshot. Also shown is the mean change in angular momentum for all particles 
inside the compact disc, which we define as all material with a semi-major
axis smaller than 1pc. In this case the values for both the isothermal and the full cooling simulations only differ after
the third decimal after the comma, hence we only show them for one case.
\label{icool_angular_tab}}
\end{table}

In order to derive the mass fraction which is bound to the black hole, we calculate the orbital structure of the gas
by using eccentricity 
\begin{equation}
\label{eccdef}
e^2 = 1 + \frac{2 J^2 E_{\rm mech}}{G^2 M_{\rm BH}^2 m^3} 
\end{equation}
where J is the angular momentum of the SPH
particle, M$_{\rm BH}$ the black hole mass, m the SPH particle mass and E$_{\rm mech}$ the total mechanical energy of the SPH particle
(E$_{\rm mech}$ = E$_{\rm potential}$ + E$_{kinetic}$).
Here eccentricity e = 0 corresponds to a circular orbit, 0 $<$ e $<$ 1 to a elliptic orbit, e = 1 to a parabolic 
or radial orbit and e $>$ 1 to a hyperbolic orbit. SPH particles with e $<$ 1 are on bound orbits, SPH particles with
e $>$ 1 are on unbound orbits and can escape the black hole potential. For SPH particles with e = 1 which are
on radial orbits, the SPH particle velocity must be compared to the black hole escape
velocity v$_{esc}^2 = \frac{2 G M_{BH}}{r}$
at SPH particle radius r in order to determine if they are bound to the black hole or if they can escape to infinity.
Table \ref{icool_bound_tab} summarises the total bound mass for all initial conditions in the full cooling and the isothermal cases.

We use initial condition V03 as an example here (cloud velocity of 50 km/s and an impact parameter of 3 pc)
since we used V03 for our numerical tests
shown in the Appendix where we performed a simulation with numerical viscosity turned off and we need this special 
setup for comparison to the standard simulations in this section.
In Figure \ref{icool_ecc_a}a we show the distribution of mass onto orbits of different eccentricity.
The black (double-dotted) line shows what we expect from initial condition V03. The red (solid) line shows the result of the
isothermal simulation using initial condition V03 with artificial viscosity turned off.
In this case (isothermal and no artificial viscosity) 
there are no large pressure gradients during the simulation so that the change in mechanical energy 
and eccentricity is small. Because of this the final distribution approximately reflects the initial conditions.
Running the same simulation (isothermal and initial condition V03) with artificial viscosity turned on leads
to the distribution shown by the blue (dotted) curve. Here it can clearly be seen that more material is on bound (e$<$1),
more circular orbits than expected from the initial conditions.

Figure \ref{icool_ecc_a}b shows the distribution of (bound) mass onto orbits of different semi-major axes for initial condition V03 in the
full cooling and isothermal case at the end of the simulations. Again this plot shows that at the end of the simulations we accumulate a
significant amount of mass onto orbits close to the black hole compared to the initial condition.
Using the information from Figure \ref{icool_ecc_a}a and \ref{icool_ecc_a}b we can calculate the degree to which the cancellation
of angular momentum contributes to the formation of the compact disc. We define the compact accretion disc to be all material with
a semi-major axis below 1pc. First we calculate the mean change in angular momentum compared to the initial angular momentum for all
particles present at the final timestep.
This is done by taking the ratio of the absolute value of the final angular momentum to the absolute value of the initial angular momentum
for every particle from the last snapshot and averaging over those values.
We ignore all particles which have already been accreted (around 2\%-15\% of the
total initial particle number) since they are no longer present in the final snapshot.
In general the mean change in angular
momentum is around 10\%-30\% of the initial angular momentum for all particles that have not been accreted. The mean values for the individual
simulations are listed in Table \ref{icool_angular_tab}. The small impact parameter simulations I01, I02 and C01 lead to a larger change
in final angular momentum compared to the large impact parameter simulations V01, V02 and V03 because of the larger amount of
material which can collide with a small impact parameter.

Cancellation of angular momentum is only strong in those parts of the cloud
streaming around the black hole from opposite sides. Since the impact parameter is not zero the cancellation of angular momentum will not
be 100\% effective and there is always a part of the cloud which just follows its initial orbit around the black hole without encountering 
material with opposite angular momentum. This material can only change its mechanical energy. Since angular momentum is conserved
this gas can only move to a larger radius. The only way to transport material below the
initial lower boundary of the semi-major axis (which is roughly 2pc in Figure \ref{icool_ecc_a}b) is the cancellation/redistribution
of angular momentum. To estimate the degree of cancellation of angular momentum needed to form the compact accretion disc we
calculate the mean change in angular momentum for material which has a semi-major axis below 1pc in the final snapshot.
The results for all simulations are summarised in the third column in Table \ref{icool_angular_tab}. Compared to the value for all particles the
cancellation needed to form the compact disc is always higher, as expected. In the extreme case of C01 more than 60\% of the initial
angular momentum needs to be cancelled in order to form the compact disc. 

Figure \ref{icool_ecc_b} shows the distribution of mass onto orbits of different eccentricity for all simulations
in the isothermal and cooling
case compared to the expected distribution from the initial conditions. As can be seen, the simulations with cooling tend 
in general to have more gas mass on more circular orbits compared to the isothermal case.
Since the orbits become more circular the smaller the distance from the black hole gets, this corresponds to more
mass closer to the black hole. In the cooling case the discs are a lot more extended in z-direction compared to the
isothermal discs (this will be discussed in detail in section \ref{icool_disc})
which basically just consist of a single sheet of SPH particles in the xy-plane. Thus, in the cooling case, material
can move a lot easier in radial direction compared to the isothermal case. In addition, in the isothermal case the
SPH particles get very close to the softening length already at a moderate distance from the black hole.
This increases smoothing over the forces and also prevents radial movement of particles.
Still the isothermal case already yields a useful approximation of the gas mass distribution compared to the
more realistic cooling case.

When calculating the initial orbits, we can also calculate the minimum distance r$_{\rm min}$ for every SPH
particle to the black hole on its specific orbit. All SPH particles which are initially on an orbit with a
distance r$_{\rm min}$ smaller than the accretion radius r$_{\rm acc}$ should be accreted. 
In Table \ref{icool_ecc_tab} we compare the accreted mass expected from this calculation to the actual accreted mass
during the whole simulation.
The simulations result in values that are a factor of 3-30 larger than naively expected from the initial orbits.
Again we see that radial movement is suppressed in the isothermal case, leading to lower values for the total
accreted mass $M_{\rm acc, iso}$ at the end of all simulations compared to the full cooling case. 
Ignoring the gas physics (taking the isothermal simulation using V03 without artificial viscosity again) we accreted
almost exactly what we would expect from initial conditions when comparing the 0.49 $\times$ 10$^3$ M$_{\sun}$ accreted
at the end of the simulation to the 0.5 $\times$ 10$^3$ M$_{\sun}$ expected just from the initial conditions.
Thus the results in table \ref{icool_ecc_tab} support the claim of our model that we can efficiently
dissipate kinetic energy and redistribute angular momentum forcing more gas onto lower, more circular orbits, often
even below the accretion radius.


\subsection{Accretion}
\label{icool_accretion}

\begin{figure*}
\begin{center}
\includegraphics[width=18cm]{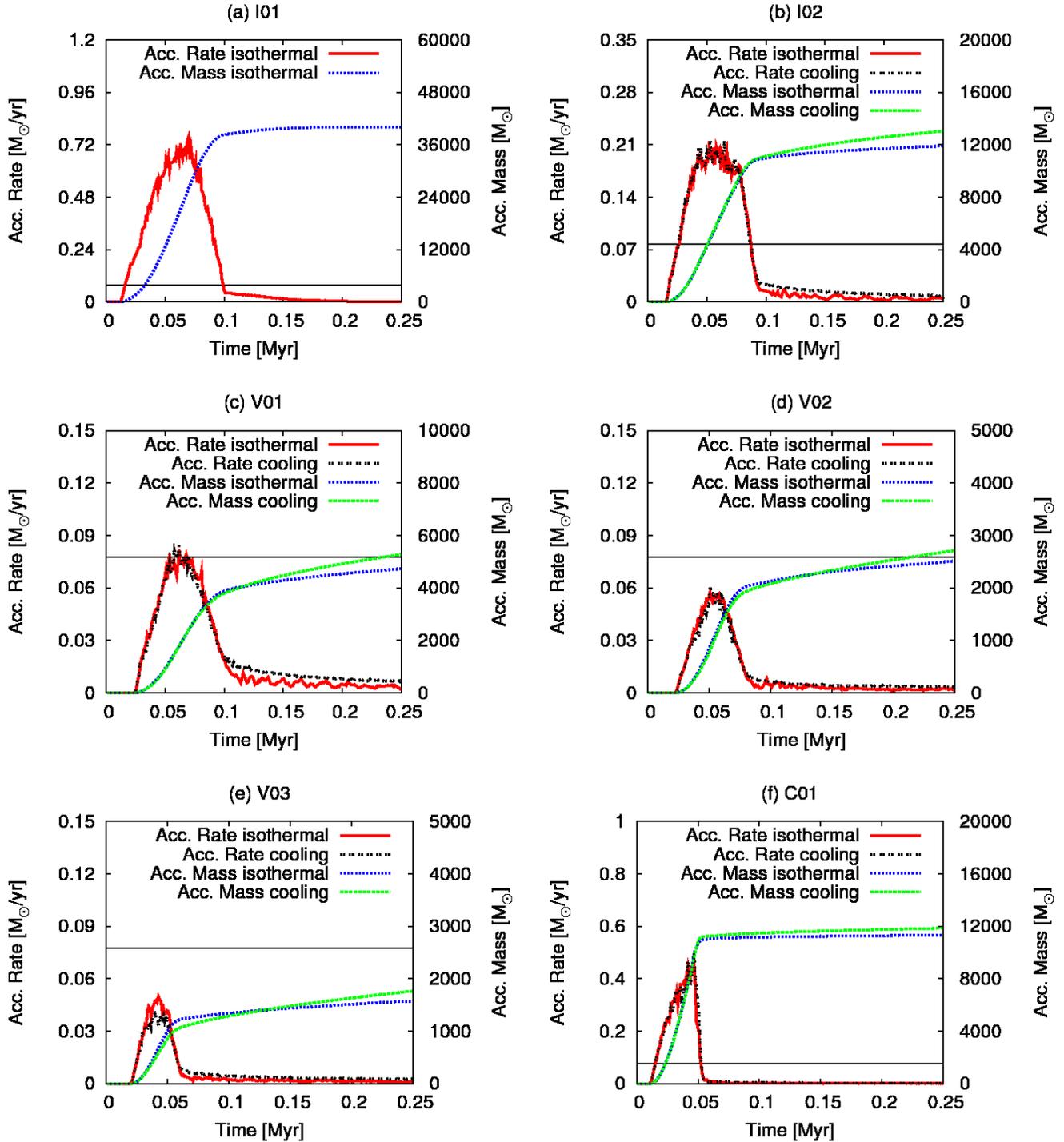}
\end{center}
\caption{The red (solid) and green (dashed) lines show the accretion rate in $\frac{M_{\sun}}{\mathrm{yr}}$ (left y-axis)
for the isothermal and cooling case.
The thin horizontal black line shows the Eddington limit of a $3.5\times 10^6$ M$_{\sun}$ black hole at 10\% accretion 
efficiency. Also plotted is the total accreted mass in blue (dotted) and black (dash-dotted) in M$_{\sun}$ (right y-axis)
again for the isothermal and full cooling cases. 
The low impact parameter simulations can accrete at Super-Eddington rates since we do not include black hole feedback,
whereas the high impact parameter simulations stay below Eddington accretion at all times.
In general the isothermal and cooling simulations show a very similar accretion behaviour.
\label{icool_acc}}
\end{figure*}

Accretion rates are calculated from material which falls below our accretion
radius of $r_{\rm acc}=2\times 10^{-2}$ pc.
We note that our simulations can lead to Super-Eddington accretion rates with the prescription we are using, since
there is no black hole feedback regulating the accretion rate. However, accretion in our case
does not necessarily mean black hole accretion since the formation of a small and hot black hole accretion
disc evolving viscously is beyond the resolution limit and outside the scope of our current simulations.
Nevertheless, it is interesting to compare our accretion rates for all simulations to the 
Eddington rate of the black hole which is shown in Figure \ref{icool_acc}. 

The Eddington mass accretion rate is defined as
$\dot{M}_{edd} = \frac{4 \pi G M_{BH} m_{p}}{\epsilon \sigma_{T} c}$, with $m_{p}$ being the proton mass,
$\epsilon$ the accretion efficiency and $\sigma_{T}$ the Thompson scattering cross-section.
From this we get an Eddington rate of $\dot{M}_{edd} = 0.0775$ $\frac{M_{\sun}}{\mathrm{yr}}$
using an accretion efficiency of 10\% and a $3.5\times 10^6$ M$_{\sun}$ black hole as in the Milky Way GC.

Figure \ref{icool_acc} shows that the isothermal and full cooling simulations differ only
marginally in their accretion behaviour, thus
we will describe accretion independent of the used equation of state. 
Note that the y-axis is scaled differently for different simulations due to the large range of
accretion rates we get.
At the end of all simulations we reach very low accretion rates indicating that we have reached a nearly 
stationary state.  The accretion rates of the large impact parameter simulations using initial conditions V02 and V03
stay below the Eddington rate at all times.

A comparison of the simulations with initial conditions I01, I02 and V02 in Table \ref{icool_ecc_tab} shows
that the impact parameter
plays the most important role in determining the final accreted mass while the cloud velocity
has a much smaller effect (compare V01, V02 and V03).
This is also confirmed by comparing the simulations using initial condition C01, which has the same initial specific
angular momentum as V03, to simulations using V03. Clearly the main factor determining the final accreted mass is
the impact parameter.
In all simulations we still have accretion at the end of the runs, however at
very low rates, thus the values shown in Table \ref{icool_ecc_tab} would only slightly increase if 
the simulations were continued for a longer time.


\subsection{Disc Structure}
\label{icool_disc}

\begin{figure*}
\begin{center}
\includegraphics[width=9cm]{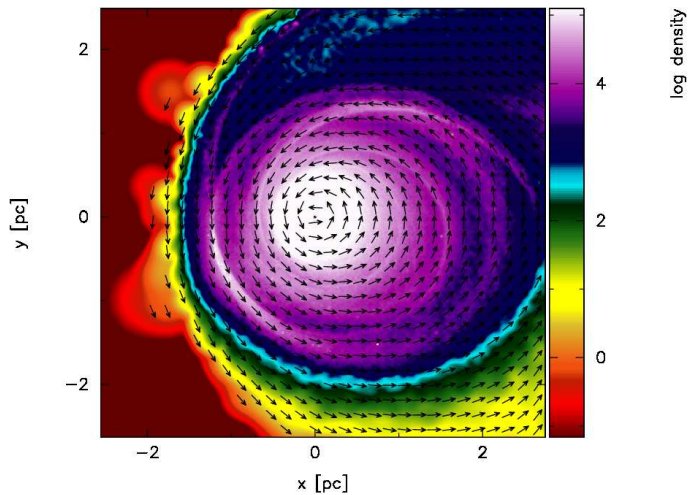}
\end{center}
\caption{
Logarithmic density in units of $\frac{M_{\sun}}{\mathrm{pc^3}}$ in the xy-plane for z=0 pc at 0.25 Myrs for
initial condition V02
with cooling. Velocity vectors are all the same length and only indicate local direction.
The semi-major axis of the final disc aligns well with the x-axis.
\label{icool_disc_a}
}
\end{figure*}

\begin{figure*}
\begin{center}
\includegraphics[width=18cm]{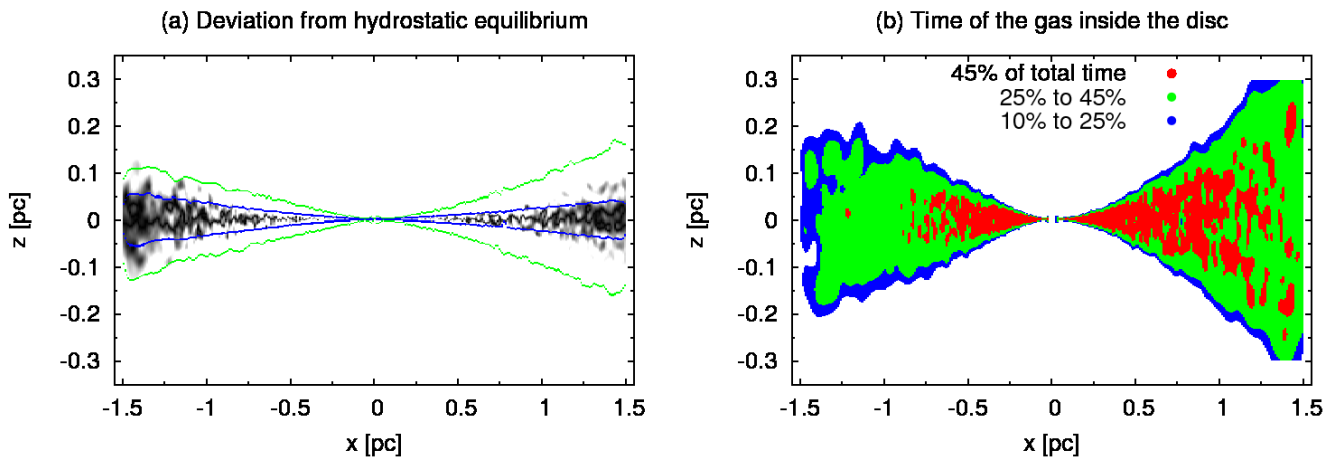}
\end{center}
\caption{
In (a) the green line shows the disc scale height resulting from our simulation for V02 in the full cooling case.
The blue line in (a) shows what is expected from theory for a hydrostatic equilibrium. The over-plotted structure in 
(a) shows the deviation of the simulation from a hydrostatic equilibrium
($\frac{\partial \phi}{\partial z} - \frac{1}{\rho} \frac{\partial P}{\partial z}$). Black indicates the lowest deviation, normalised
to the global minimum deviation. The colour gradient from black to white spans two orders of magnitude in the deviation from
the global minimum deviation. The pure-black part which is close to a hydrostatic equilibrium is in good agreement
with the expectation from theory (blue line).
Plot (b) shows the time the gas spent inside the accretion disc.
Red shows all gas
which spent more than 45\% of the total time inside the disc, green all gas which spent 25\% to 45\% of the total time
inside the disc and blue shows all gas which spent 10\% to 25\% of the total time inside the disc.
The old (red) gas settles into the disc mid-plane whereas the young (green/blue) gas builds up the atmosphere which
is not yet in hydrostatic equilibrium (comparing gas at fixed radius along the z-axis).
We also see that the disc is growing inside out since in general material at larger radii is younger.
This hides the atmosphere described above in the very inner parts of
the disc where all gas is very old and in the outer parts where all gas is very young. At around $\pm$ 0.75 pc
on the x-axis the atmosphere
around the older gas that already settled into the disc-midplane can be seen the best.
\label{icool_disc_b}
}
\end{figure*}

Finally, we study the structure of our discs in the z-direction at the final timestep. 
For this we define our disc scale height to be the position where the
density in z-direction drops to e$^{-1}$ times the mid-plane density. We take the simulation using initial 
condition V02 as an example since in this case the disc aligns well in the xy-plane with the semi-major axis
along the x-axis as can be seen in Figure \ref{icool_disc_a}.

\subsubsection{The Isothermal Case}

For a standard thin accretion disc assuming hydrostatic equilibrium in z-direction, where thermal
pressure is balanced by the z-component of the central force we can calculate the disc scale height.
The resulting disc scale height is H = $c_s \left( \frac{r_{xy}^3}{GM} \right)^{1/2}$ with r$_{xy}$
the SPH particle distance to the black hole in the xy-plane and c$_s$ the SPH particle sound speed.
For our initial temperature of 50 K in the isothermal case we get a scale height of 5$\times$10$^{-5}$ pc at
the inner boundary of 0.02 pc and a scale height of the order of the softening length at 0.5 pc.
Thus the accretion disc scale height gets artificially blown up
to our softening length of $\epsilon = 10^{-3}$ pc in the inner region. The disc scale height
should be resolved to at least four times the softening length according to \cite{2006MNRAS.373.1039N},
thus we would need a softening length of around $10^{-5}$ pc or preferably better at the inner boundary.

This is however not feasible with respect to the simulation runtime. 
From this we can conclude that it is important to include the detailed thermodynamical treatment of the full
cooling prescription in order to investigate the z-structure of the accretion disc, which is more
extended due to the higher temperature and thus can be resolved using a realistic value for the softening length. 

\subsubsection{The Full Cooling Case}

Here we resolve the disc scale height to at least four times the softening length at a radius of
0.1 pc and up to 40 times the softening length at a radius of 1 pc.
A test-simulation using initial condition V03 and a softening length of $\epsilon$ = 10$^{-4}$
(presented in the Appendix) resolving the disc scale height to one order of magnitude more in softening
length shows no difference in the results compared to the larger softening length of $\epsilon$ = 10$^{-3}$
indicating that we reached convergence.

The green line in Figure \ref{icool_disc_b}a shows the disc scale height resulting from our simulations for a y=0 pc slice along
the x-axis for the disc shown in Figure \ref{icool_disc_a} in the full cooling case.
This is compared to what we would expect from assuming hydrostatic equilibrium in z-direction,
shown by the blue line in Figure \ref{icool_disc_b}a.
The over-plotted structure shows the absolute value of $\frac{\partial \phi}{\partial z} - \frac{1}{\rho} \frac{\partial P}{\partial z}$ with $\phi$ the point-mass
potential of the black hole. Black indicates the minimum value (normalised to the global minimum value). The colour gradient from black to white spans two orders
of magnitude in the deviation from the global minimum deviation. 
We see that the central part of the disc (pure-black in the plot) is close to hydrostatic equilibrium with a good fit to the expected theoretical value (blue line).
The outer parts of the disc consist of fresh material that just fell onto the disc and did not yet settle into hydrostatic
equilibrium.

This can be seen in Figure \ref{icool_disc_b}b, where the age of the gas inside the disc is plotted. Red shows all gas
which already spent more than 45\% of the total time inside the disc, green all gas which spent 25\% to 45\% of the total time
inside the disc and blue shows all gas which spent 10\% to 25\% of the total time inside the disc.
Old (red) gas is concentrated more in the disc mid-plane, whereas fresh (green/blue) gas builds up the atmosphere of the disc, which
is not yet in hydrostatic equilibrium (comparing gas at fixed radius along the z-axis).
In addition Figure \ref{icool_disc_b}b also shows that the disc is growing inside out since 
in general the material at larger radii is younger. This hides the atmosphere described above in the very inner parts of
the disc where all gas is very old and in the outer parts where all gas is very young. At around $\pm$ 0.75 pc on the x-axis the atmosphere
around the older gas that already settled into the disc-midplane can be seen the best.


\section{Summary and Discussion}
\label{summary}

We have performed simulations of a gas cloud colliding with a super-massive
black hole, with parts of the cloud engulfing the black hole during the
process. We demonstrate that even when the cloud is initially unbound to the black hole
due to a high velocity, sub parsec-scale gas discs can form.
Clearly, there are a lot of additional details that can be studied, which we will  
explore in a subsequent paper.
Firstly, we did not investigate fragmentation and star formation yet which is necessary
to directly compare the simulations with the observations.
Secondly, the potential of the stellar cusp observed in the GC clearly influences SPH particle
orbits around the black hole.
Finally, black hole and stellar feedback could also play an important role in 
the formation process.

An adiabatic equation of state leads to an explosion of the cloud during infall
with thermal energies comparable to supernova energies of 10$^{51}$ erg.
This prevents formation of a compact accretion disc and leads to a large expanding bubble
of hot gas. For very inefficient cooling, this could be an alternative interpretation for a hot bubble
of gas seen near a black hole like e.g. Sgr A East in our milky-way GC, which has been attributed
to a supernova explosion (\citealt{2005ApJ...620..287H}).

The isothermal case produces already
viable results for the accretion behaviour and distribution of mass onto orbits of different eccentricity
compared to the full cooling case. However, numerical and other problems especially with resolving the disc in
z-direction lead to the conclusion that a full thermodynamical treatment is necessary.
In addition, the isothermal approach enhances fragmentation strongly. Due to the low Jeans-mass at
a temperature of 50 K a huge number of low mass fragments forms. To study fragmentation in detail, again
a full thermodynamical treatment will be necessary.

The simulations using the full cooling prescription of \cite{2007A&A...475...37S} represent the most
realistic case. Here we are not dominated by numerics and resolve the disc scale height to multiple
times the softening length due to the higher temperatures resulting in a larger scale height.
In this case we do not artificially suppress radial movement, leading to more circular orbits 
compared to the isothermal case.

We find that the disc mid-plane consists of old (= already inside the disc for a long time) gas close to hydrostatic equilibrium,
surrounded by an atmosphere
of young gas that was just falling in and that did not yet settle into hydrostatic equilibrium. 
This explains the deviation of the disc
scale height we find in our simulations compared to what we would expect from gas in hydrostatic equilibrium.

Generally we find that the impact parameter is the dominant factor in determining the accretion rates,
from Sub-Eddington accretion rates for large impact parameters to Super-Eddington
accretion for small impact parameters.
Cloud velocity determines the final rotation of the resulting discs semi-major axis with respect to the x-axis
in the xy-plane. Fast clouds (v$_{\rm cloud} >$ 50 km/s) lead to discs that are rotated towards the negative y-axis and slow clouds
(v$_{\rm cloud} <$ 50 km/s) to discs that are rotated towards the positive y-axis.
Also cloud velocity strongly determines the amount of gas bound to the black hole initially. Low velocities lead to more
mass inside the final disc compared to simulations with velocities such that nearly no SPH particle is
bound to the black hole initially.

The goal of this paper is to provide a proof of principle of the idea of a cloud
engulfing the black hole during passage. With this mechanism we are able to produce parsec 
to sub-parsec sized dense accretion discs even when the initial cloud is unbound to the black
hole.


\section*{Acknowledgements}

We would like to thank the Reviewer for very useful comments and constructive 
criticism, that helped us to improve the paper significantly.
We would like to thank Jim Pringle for providing useful comments on the paper.
This research was supported by the DFG cluster of excellence "Origin and Structure of the Universe".
Simulations were performed on an SGI Altix 3700 Bx2 supercomputer partly funded by the DFG cluster of
excellence "Origin and Structure of the Universe" and the HLRBII supercomputer at the LRZ M\"unchen.
Most of the plots have been created using the publicly available SPH visualisation tool SPLASH
by D.J. Price \citep{2007PASA...24..159P}.

\bibliographystyle{mn2e}
\bibliography{ref}

\begin{thebibliography}{43}
\expandafter\ifx\csname natexlab\endcsname\relax\def\natexlab#1{#1}\fi

\bibitem[{{Agertz} {et~al.}(2007){Agertz}, {Moore}, {Stadel}, {Potter},
  {Miniati}, {Read}, {Mayer}, {Gawryszczak}, {Kravtsov}, {Nordlund}, {Pearce},
  {Quilis}, {Rudd}, {Springel}, {Stone}, {Tasker}, {Teyssier}, {Wadsley}, \&
  {Walder}}]{2007MNRAS.380..963A}
{Agertz} O., {Moore} B., {Stadel} J., {Potter} D., {Miniati} F., {Read} J.,
  {Mayer} L., {Gawryszczak} A., {Kravtsov} A., {Nordlund} {\AA}., {Pearce} F.,
  {Quilis} V., {Rudd} D., {Springel} V., {Stone} J., {Tasker} E., {Teyssier}
  R., {Wadsley} J., {Walder} R., 2007, \mnras, 380, 963

\bibitem[{{Alexander} {et~al.}(2008){Alexander}, {Armitage}, {Cuadra}, \&
  {Begelman}}]{2008ApJ...674..927A}
{Alexander} R.~D., {Armitage} P.~J., {Cuadra} J., {Begelman} M.~C., 2008, \apj,
  674, 927

\bibitem[{{Alexander} {et~al.}(2007){Alexander}, {Begelman}, \&
  {Armitage}}]{2007ApJ...654..907A}
{Alexander} R.~D., {Begelman} M.~C., {Armitage} P.~J., 2007, \apj, 654, 907

\bibitem[{{Balsara}(1995)}]{1995JCoPh.121..357B}
{Balsara} D.~S., 1995, Journal of Computational Physics, 121, 357

\bibitem[{{Bartko} {et~al.}(2009){Bartko}, {Martins}, {Fritz}, {Genzel},
  {Levin}, {Perets}, {Paumard}, {Nayakshin}, {Gerhard}, {Alexander},
  {Dodds-Eden}, {Eisenhauer}, {Gillessen}, {Mascetti}, {Ott}, {Perrin},
  {Pfuhl}, {Reid}, {Rouan}, {Sternberg}, \& {Trippe}}]{2009ApJ...697.1741B}
{Bartko} H., {Martins} F., {Fritz} T.~K., {Genzel} R., {Levin} Y., {Perets}
  H.~B., {Paumard} T., {Nayakshin} S., {Gerhard} O., {Alexander} T.,
  {Dodds-Eden} K., {Eisenhauer} F., {Gillessen} S., {Mascetti} L., {Ott} T.,
  {Perrin} G., {Pfuhl} O., {Reid} M.~J., {Rouan} D., {Sternberg} A., {Trippe}
  S., 2009, \apj, 697, 1741

\bibitem[{{Bartko} {et~al.}(2010){Bartko}, {Martins}, {Trippe}, {Fritz},
  {Genzel}, {Ott}, {Eisenhauer}, {Gillessen}, {Paumard}, {Alexander},
  {Dodds-Eden}, {Gerhard}, {Levin}, {Mascetti}, {Nayakshin}, {Perets},
  {Perrin}, {Pfuhl}, {Reid}, {Rouan}, {Zilka}, \&
  {Sternberg}}]{2010ApJ...708..834B}
{Bartko} H., {Martins} F., {Trippe} S., {Fritz} T.~K., {Genzel} R., {Ott} T.,
  {Eisenhauer} F., {Gillessen} S., {Paumard} T., {Alexander} T., {Dodds-Eden}
  K., {Gerhard} O., {Levin} Y., {Mascetti} L., {Nayakshin} S., {Perets} H.~B.,
  {Perrin} G., {Pfuhl} O., {Reid} M.~J., {Rouan} D., {Zilka} M., {Sternberg}
  A., 2010, \apj, 708, 834

\bibitem[{{Bate} \& {Burkert}(1997)}]{1997MNRAS.288.1060B}
{Bate} M.~R., {Burkert} A., 1997, \mnras, 288, 1060

\bibitem[{{Bonnell} \& {Rice}(2008)}]{2008Sci...321.1060B}
{Bonnell} I.~A., {Rice} W.~K.~M., 2008, Science, 321, 1060

\bibitem[{{Cuadra} {et~al.}(2008){Cuadra}, {Armitage}, \&
  {Alexander}}]{2008MNRAS.388L..64C}
{Cuadra} J., {Armitage} P.~J., {Alexander} R.~D., 2008, \mnras, 388, L64

\bibitem[{{Genzel} {et~al.}(2003){Genzel}, {Sch{\"o}del}, {Ott}, {Eisenhauer},
  {Hofmann}, {Lehnert}, {Eckart}, {Alexander}, {Sternberg}, {Lenzen},
  {Cl{\'e}net}, {Lacombe}, {Rouan}, {Renzini}, \&
  {Tacconi-Garman}}]{2003ApJ...594..812G}
{Genzel} R., {Sch{\"o}del} R., {Ott} T., {Eisenhauer} F., {Hofmann} R.,
  {Lehnert} M., {Eckart} A., {Alexander} T., {Sternberg} A., {Lenzen} R.,
  {Cl{\'e}net} Y., {Lacombe} F., {Rouan} D., {Renzini} A., {Tacconi-Garman}
  L.~E., 2003, \apj, 594, 812

\bibitem[{{Ghez} {et~al.}(2005){Ghez}, {Salim}, {Hornstein}, {Tanner}, {Lu},
  {Morris}, {Becklin}, \& {Duch{\^e}ne}}]{2005ApJ...620..744G}
{Ghez} A.~M., {Salim} S., {Hornstein} S.~D., {Tanner} A., {Lu} J.~R., {Morris}
  M., {Becklin} E.~E., {Duch{\^e}ne} G., 2005, \apj, 620, 744

\bibitem[{{Gingold} \& {Monaghan}(1983)}]{1983MNRAS.204..715G}
{Gingold} R.~A., {Monaghan} J.~J., 1983, \mnras, 204, 715

\bibitem[{{G{\"u}sten} \& {Philipp}(2004)}]{2004dimg.conf..253G}
{G{\"u}sten} R., {Philipp} S.~D., 2004, in The Dense Interstellar Medium in
  Galaxies, {Pfalzner} S., {Kramer} C., {Staubmeier} C., {Heithausen} A., eds.,
  p. 253

\bibitem[{{Herrnstein} \& {Ho}(2005)}]{2005ApJ...620..287H}
{Herrnstein} R.~M., {Ho} P.~T.~P., 2005, \apj, 620, 287

\bibitem[{{Hobbs} \& {Nayakshin}(2009)}]{2009MNRAS.394..191H}
{Hobbs} A., {Nayakshin} S., 2009, \mnras, 394, 191

\bibitem[{{Ivanov} {et~al.}(2005){Ivanov}, {Polnarev}, \&
  {Saha}}]{2005MNRAS.358.1361I}
{Ivanov} P.~B., {Polnarev} A.~G., {Saha} P., 2005, \mnras, 358, 1361

\bibitem[{{Johansson} {et~al.}(2009){Johansson}, {Naab}, \&
  {Burkert}}]{Johansson:2009BH}
{Johansson} P.~H., {Naab} T., {Burkert} A., 2009, \apj, 690, 802

\bibitem[{{King} \& {Pringle}(2007)}]{2007MNRAS.377L..25K}
{King} A.~R., {Pringle} J.~E., 2007, \mnras, 377, L25

\bibitem[{{Kolykhalov} \& {Syunyaev}(1980)}]{1980SvAL....6..357K}
{Kolykhalov} P.~I., {Syunyaev} R.~A., 1980, Soviet Astronomy Letters, 6, 357

\bibitem[{{Levin} \& {Beloborodov}(2003)}]{2003ApJ...590L..33L}
{Levin} Y., {Beloborodov} A.~M., 2003, \apjl, 590, L33

\bibitem[{{L{\"o}ckmann} {et~al.}(2009){L{\"o}ckmann}, {Baumgardt}, \&
  {Kroupa}}]{2009MNRAS.398..429L}
{L{\"o}ckmann} U., {Baumgardt} H., {Kroupa} P., 2009, \mnras, 398, 429

\bibitem[{{Lodato} \& {Rice}(2004)}]{2004MNRAS.351..630L}
{Lodato} G., {Rice} W.~K.~M., 2004, \mnras, 351, 630

\bibitem[{{Lu} {et~al.}(2009){Lu}, {Ghez}, {Hornstein}, {Morris}, {Becklin}, \&
  {Matthews}}]{2009ApJ...690.1463L}
{Lu} J.~R., {Ghez} A.~M., {Hornstein} S.~D., {Morris} M.~R., {Becklin} E.~E.,
  {Matthews} K., 2009, \apj, 690, 1463

\bibitem[{{Madigan} {et~al.}(2009){Madigan}, {Levin}, \&
  {Hopman}}]{2009ApJ...697L..44M}
{Madigan} A.-M., {Levin} Y., {Hopman} C., 2009, \apjl, 697, L44

\bibitem[{{Mapelli} {et~al.}(2008){Mapelli}, {Hayfield}, {Mayer}, \&
  {Wadsley}}]{2008arXiv0805.0185M}
{Mapelli} M., {Hayfield} T., {Mayer} L., {Wadsley} J., 2008, ArXiv e-prints,
  arXiv:0805.0185

\bibitem[{{Miyazaki} \& {Tsuboi}(2000)}]{2000ApJ...536..357M}
{Miyazaki} A., {Tsuboi} M., 2000, \apj, 536, 357

\bibitem[{{Montero-Casta{\~n}o} {et~al.}(2009){Montero-Casta{\~n}o},
  {Herrnstein}, \& {Ho}}]{2009ApJ...695.1477M}
{Montero-Casta{\~n}o} M., {Herrnstein} R.~M., {Ho} P.~T.~P., 2009, \apj, 695,
  1477

\bibitem[{{Murray}(1996)}]{1996MNRAS.279..402M}
{Murray} J.~R., 1996, \mnras, 279, 402

\bibitem[{{Nayakshin} \& {Cuadra}(2005)}]{2005A&A...437..437N}
{Nayakshin} S., {Cuadra} J., 2005, \aap, 437, 437

\bibitem[{{Nayakshin} {et~al.}(2006){Nayakshin}, {Dehnen}, {Cuadra}, \&
  {Genzel}}]{2006MNRAS.366.1410N}
{Nayakshin} S., {Dehnen} W., {Cuadra} J., {Genzel} R., 2006, \mnras, 366, 1410

\bibitem[{{Nelson}(2006)}]{2006MNRAS.373.1039N}
{Nelson} A.~F., 2006, \mnras, 373, 1039

\bibitem[{{Paumard} {et~al.}(2006){Paumard}, {Genzel}, {Martins}, {Nayakshin},
  {Beloborodov}, {Levin}, {Trippe}, {Eisenhauer}, {Ott}, {Gillessen}, {Abuter},
  {Cuadra}, {Alexander}, \& {Sternberg}}]{2006ApJ...643.1011P}
{Paumard} T., {Genzel} R., {Martins} F., {Nayakshin} S., {Beloborodov} A.~M.,
  {Levin} Y., {Trippe} S., {Eisenhauer} F., {Ott} T., {Gillessen} S., {Abuter}
  R., {Cuadra} J., {Alexander} T., {Sternberg} A., 2006, \apj, 643, 1011

\bibitem[{{Poliachenko} \& {Shukhman}(1977)}]{1977SvAL....3..134P}
{Poliachenko} V.~L., {Shukhman} I.~G., 1977, Soviet Astronomy Letters, 3, 134

\bibitem[{{Price}(2007)}]{2007PASA...24..159P}
{Price} D.~J., 2007, Publications of the Astronomical Society of Australia, 24,
  159

\bibitem[{{Rice} {et~al.}(2005){Rice}, {Lodato}, \&
  {Armitage}}]{2005MNRAS.364L..56R}
{Rice} W.~K.~M., {Lodato} G., {Armitage} P.~J., 2005, \mnras, 364, L56

\bibitem[{{Sanders}(1998)}]{1998MNRAS.294...35S}
{Sanders} R.~H., 1998, \mnras, 294, 35

\bibitem[{{Shlosman} \& {Begelman}(1987)}]{1987Natur.329..810S}
{Shlosman} I., {Begelman} M.~C., 1987, \nat, 329, 810

\bibitem[{{Springel}(2005)}]{2005MNRAS.364.1105S}
{Springel} V., 2005, \mnras, 364, 1105

\bibitem[{{Stamatellos} {et~al.}(2007){Stamatellos}, {Whitworth}, {Bisbas}, \&
  {Goodwin}}]{2007A&A...475...37S}
{Stamatellos} D., {Whitworth} A.~P., {Bisbas} T., {Goodwin} S., 2007, \aap,
  475, 37

\bibitem[{{Steinmetz} \& {Mueller}(1993)}]{1993A&A...268..391S}
{Steinmetz} M., {Mueller} E., 1993, \aap, 268, 391

\bibitem[{{Ulubay-Siddiki} {et~al.}(2009){Ulubay-Siddiki}, {Gerhard}, \&
  {Arnaboldi}}]{2009MNRAS.398..535U}
{Ulubay-Siddiki} A., {Gerhard} O., {Arnaboldi} M., 2009, \mnras, 398, 535

\bibitem[{{Vollmer} {et~al.}(2003){Vollmer}, {Duschl}, \&
  {Zylka}}]{2003ANS...324..613V}
{Vollmer} B., {Duschl} W.~J., {Zylka} R., 2003, Astronomische Nachrichten
  Supplement, 324, 613

\bibitem[{{Wardle} \& {Yusef-Zadeh}(2008)}]{2008ApJ...683L..37W}
{Wardle} M., {Yusef-Zadeh} F., 2008, \apjl, 683, L37

\end{thebibliography}

\newpage
\ \\


\begin{appendix}

\section{Numerical Stability}
\label{numstab}

In this section we present results of our numerical stability test-simulations, for which we used isothermal
and full cooling simulations with initial condition V03.

\subsection{Artificial Viscosity}
\label{numstab_alpha}

\begin{figure*}
\begin{center}
\includegraphics[width=18cm]{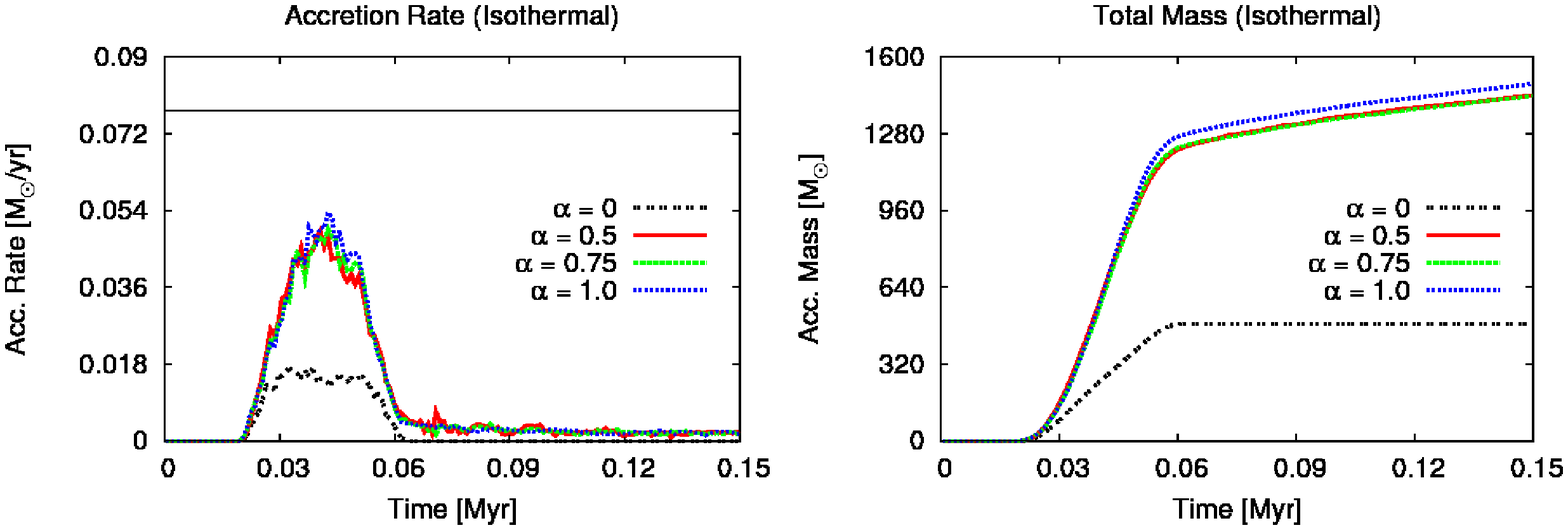}
\end{center}
\caption{The impact on accretion rate and total accreted mass as a function of the numerical
viscosity $\alpha$ parameter for the isothermal case. To visualise the differences better, the plots are shown
only from 0 to 0.15 Myrs.
Red (solid) shows the results for $\alpha$ = 0.5, green (dashed) shows the results for $\alpha$ = 0.75
and blue (dotted) the results for $\alpha$ = 1.0. The black line (dashed-dotted) shows the special 
case $\alpha$ = 0, which turns off the interaction of gas so that only material which initially already
was on an orbit smaller than the accretion radius gets accreted. 
As expected with higher viscosity the gas interacts stronger and thus more material falls onto the black hole,
but the results are not strongly dominated by the actual choice of the viscosity parameter. 
The thin black horizontal line in the accretion plot again shows the Eddington limit for the GC black hole.
\label{numstab_vis_a}}
\end{figure*}

\begin{figure*}
\begin{center}
\includegraphics[width=18cm]{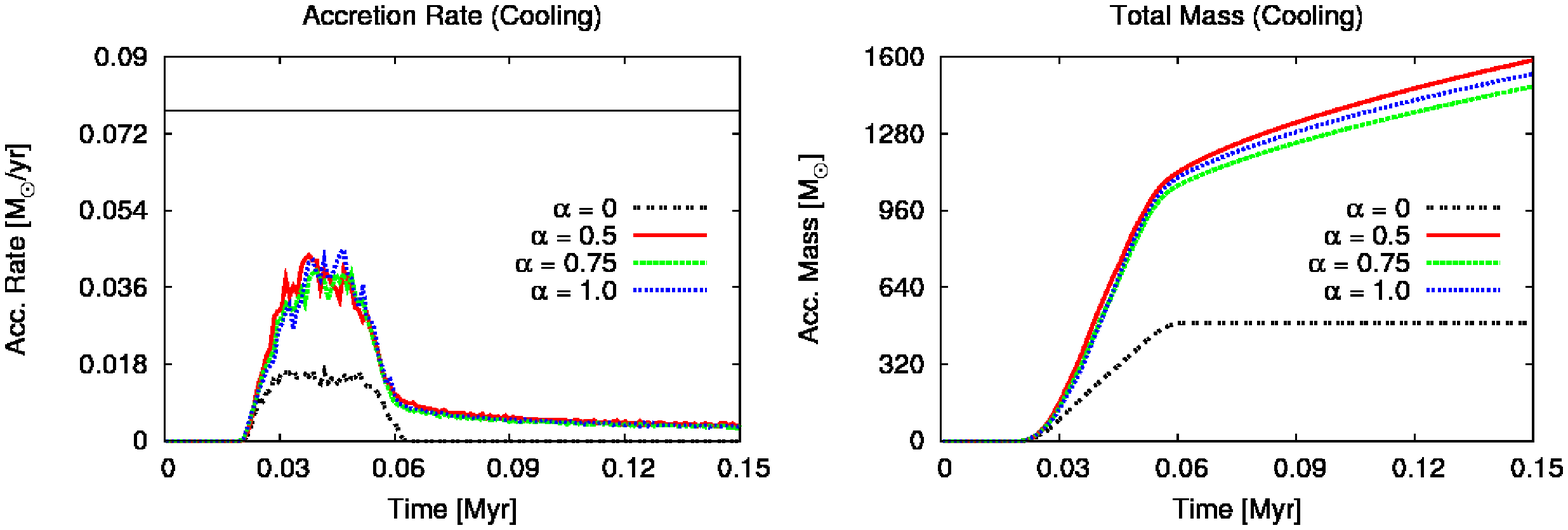}
\end{center}
\caption{Same plot as \ref{numstab_vis_a}, but for the cooling case.
\label{numstab_vis_b}}
\end{figure*}

\begin{figure*}
\begin{center}
\includegraphics[width=18cm]{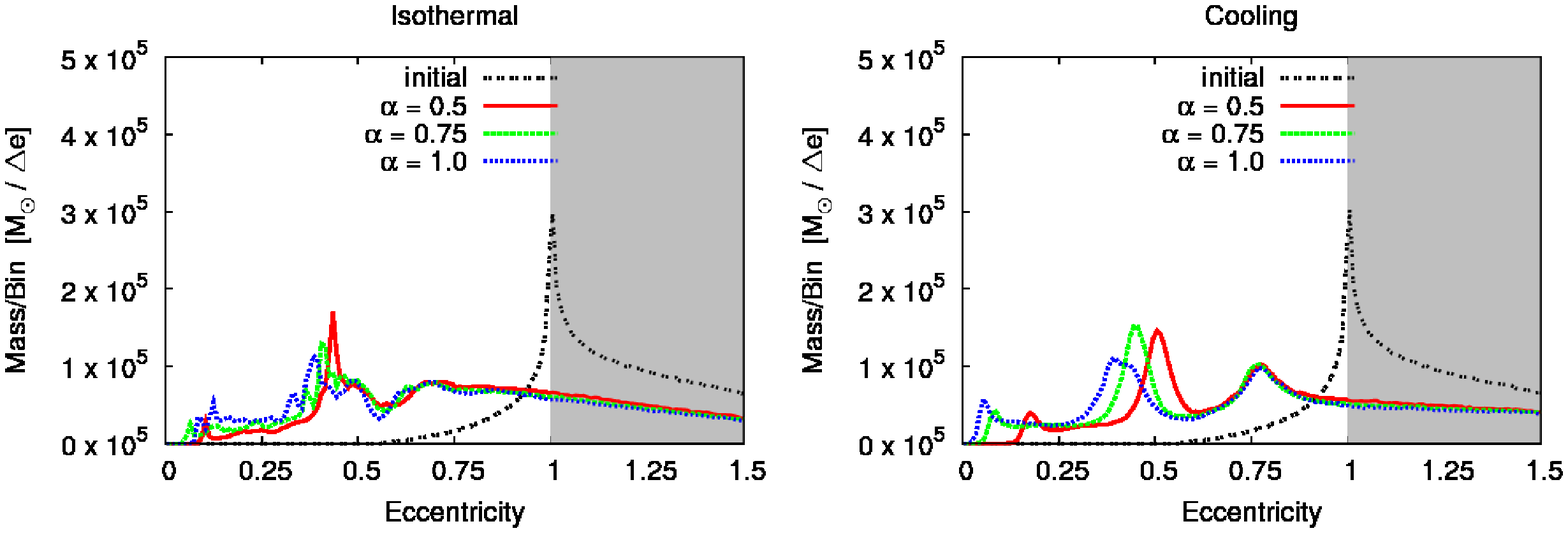}
\end{center}
\caption{Comparison of the distribution of mass onto orbits of different eccentricity depending on numerical
viscosity $\alpha$ for the isothermal and cooling case. 
SPH particles with e $>$ 1 are not bound to the black hole, indicated by the gray background.
In the isothermal case the suppression of radial movement as described in section \ref{icool_ecc} leads to
a similar behaviour for all values of the $\alpha$ parameter. In the full cooling case we expect 
a larger difference in the results due to stronger interaction of the initial colliding flows of gas with a larger
$\alpha$ parameter and the relative ease of radial movement because of the higher disc scale-height.
The choice of $\alpha$ = 0.75 gives a good mean between the results of the suggested range of $0.5 < \alpha < 1$. 
\label{numstab_vis_c}}
\end{figure*}

To study the dependence of our results on artificial viscosity we performed a parameter-study in $\alpha$ of the
artificial viscosity presented in section \ref{num} for the isothermal and the full cooling case.
The range of $\alpha$ suited for simulations is $\alpha$ = 0.5 - 1.0
\citep{2005MNRAS.364.1105S}. We performed test simulations with $\alpha$ = 0, 0.5 and 1.0. The simulations presented
throughout the paper used $\alpha$ = 0.75.
Figure \ref{numstab_vis_a} and \ref{numstab_vis_b} show the dependence of accretion rate and total accreted mass on
the $\alpha$ parameter in the isothermal and cooling case.
As can be seen, the results are stable as a function of $\alpha$ within the range suited for simulations.
The special case $\alpha$ = 0 leads to no transport of material at all into the accretion boundary region,
only material which initially already was on an orbit smaller than the accretion radius gets accreted here.
As expected, a high viscosity parameter leads to a somewhat higher transport of material, however
this artificially enhanced gas transport is negligible compared to the physical processes described in this paper.

Figure \ref{numstab_vis_c} shows the distribution of mass onto orbits of different eccentricity 
for different values of the numerical viscosity parameter $\alpha$.
As described in section \ref{icool_ecc} in the isothermal case radial movement suppression gets larger the closer
we get to the black hole. This leads to the difference especially at low eccentricities between the full cooling case
and the isothermal case. In the full cooling case different values for the strength of the numerical viscosity lead
to a stronger effect on the final distribution compared to the isothermal case, as material can
move in radial direction more easily. We do not expect the resulting distributions to be exactly the same for the cooling case
since clearly a stronger viscous force should lead to more interaction of the initial colliding flows of gas, transporting
more material onto lower, more circular orbits. However, we are not dominated by this effect, as the outer parts, the total
amount of gas and the general shape are very similar for all cases.
The choice of $\alpha$ = 0.75 gives a good mean between the results of the suggested range of $0.5 < \alpha < 1$.


\subsection{Resolution}
\label{numstab_res}

\begin{figure*}
\begin{center}
\includegraphics[width=18cm]{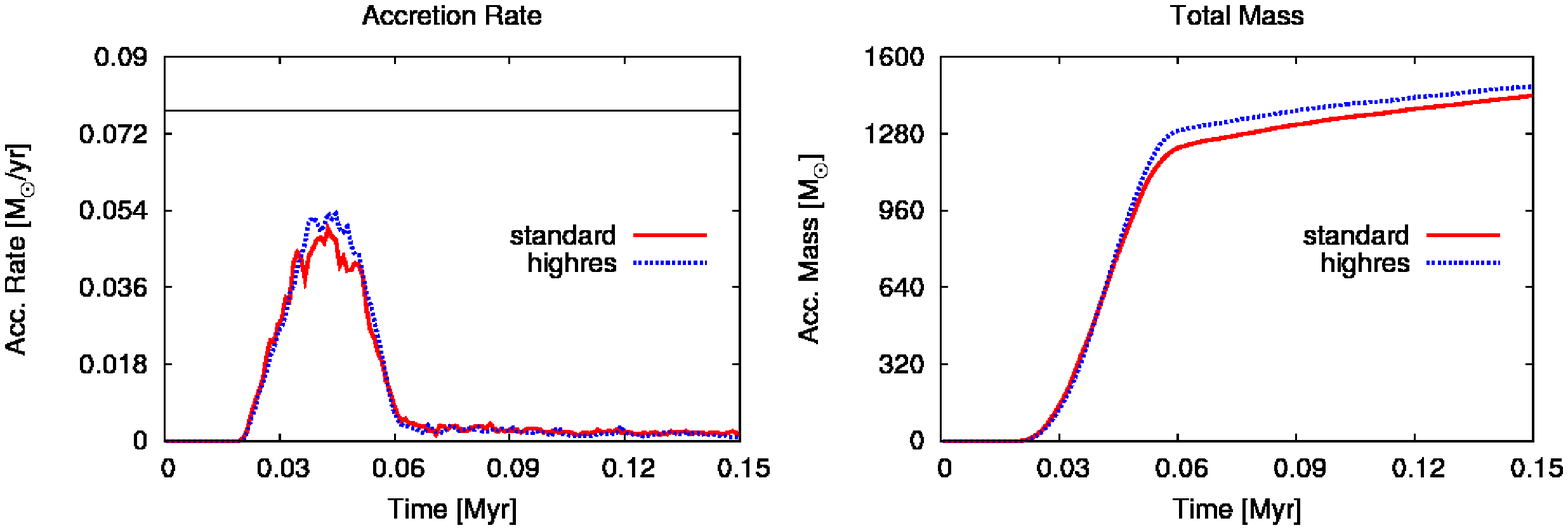}
\end{center}
\caption{
Impact on accretion rate and total accreted mass as a function of simulation resolution in 
the isothermal case. Red (solid) shows
the results for 10$^6$ SPH particles, blue (dotted) the results for 5$\times$10$^6$ SPH particles.
To visualise the differences better, the plots are shown again 
only from 0 to 0.15 Myrs.
The high resolution simulation shows somewhat larger accretion rates during the impact.
\label{numstab_res_a}}
\end{figure*}

\begin{figure*}
\begin{center}
\includegraphics[width=9cm]{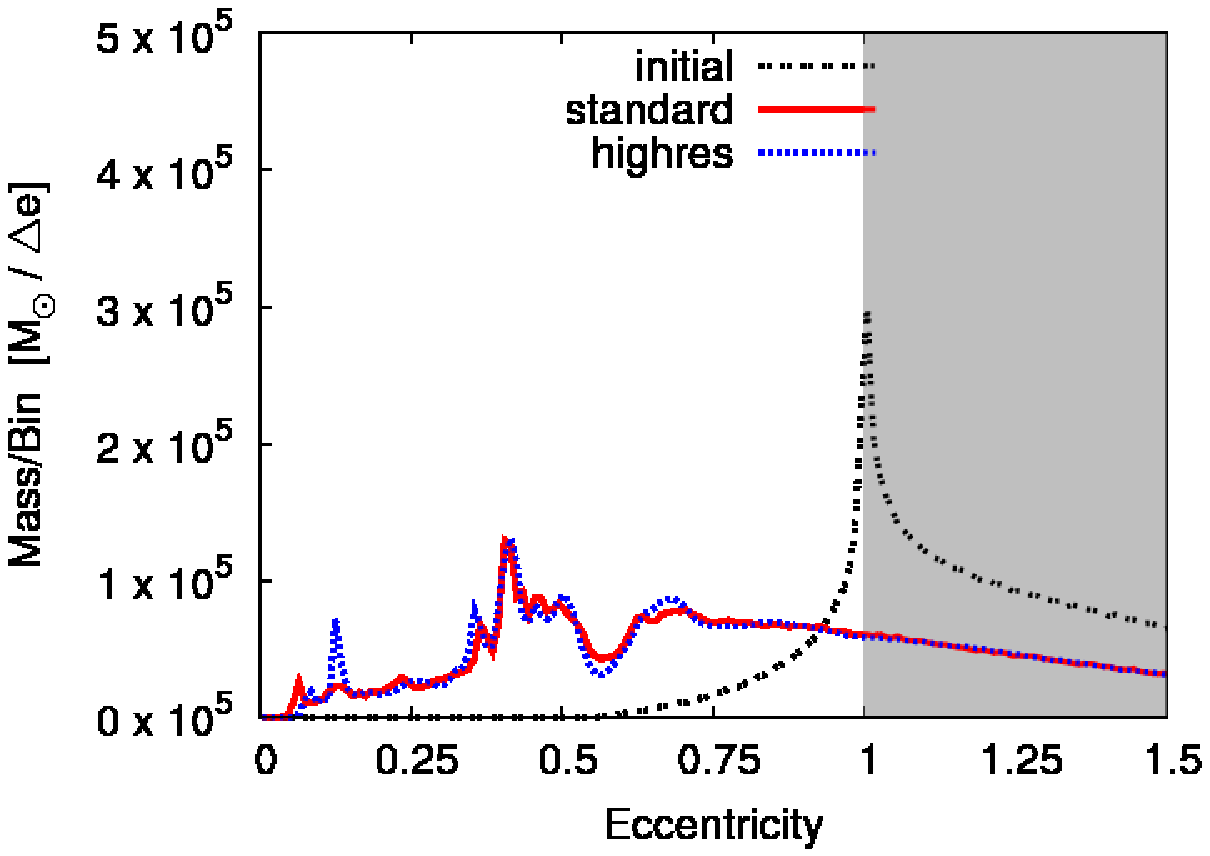}
\end{center}
\caption{Comparison of the distribution of mass onto orbits of different eccentricity for our
standard resolution and a high-resolution test. The difference
is very small, with the high resolution simulation leading to a bit more circular orbits compared to the lower resolution case.
SPH particles with e $>$ 1 are not bound to the black hole, indicated by the grey background.
\label{numstab_res_b}}
\end{figure*}

To study the effect of resolution on our results we repeated the isothermal simulation using
V03 with SPH particle number increased to
5$\times$10$^6$ SPH particles, compared to 10$^6$ SPH particles for the normal runs.
Figure \ref{numstab_res_a} shows the resulting accretion rates and total accreted mass.
The high resolution simulation shows slightly stronger accretion at the beginning.

Figure \ref{numstab_res_b} shows the distribution of mass onto orbits of different eccentricity for the high and standard
resolution isothermal simulation. Here the difference is very small with the high resolution simulation leading to slightly
more circular orbits compared to the lower resolution case.


\subsection{Softening Length}
\label{numstab_soft}

\begin{figure*}
\begin{center}
\includegraphics[width=18cm]{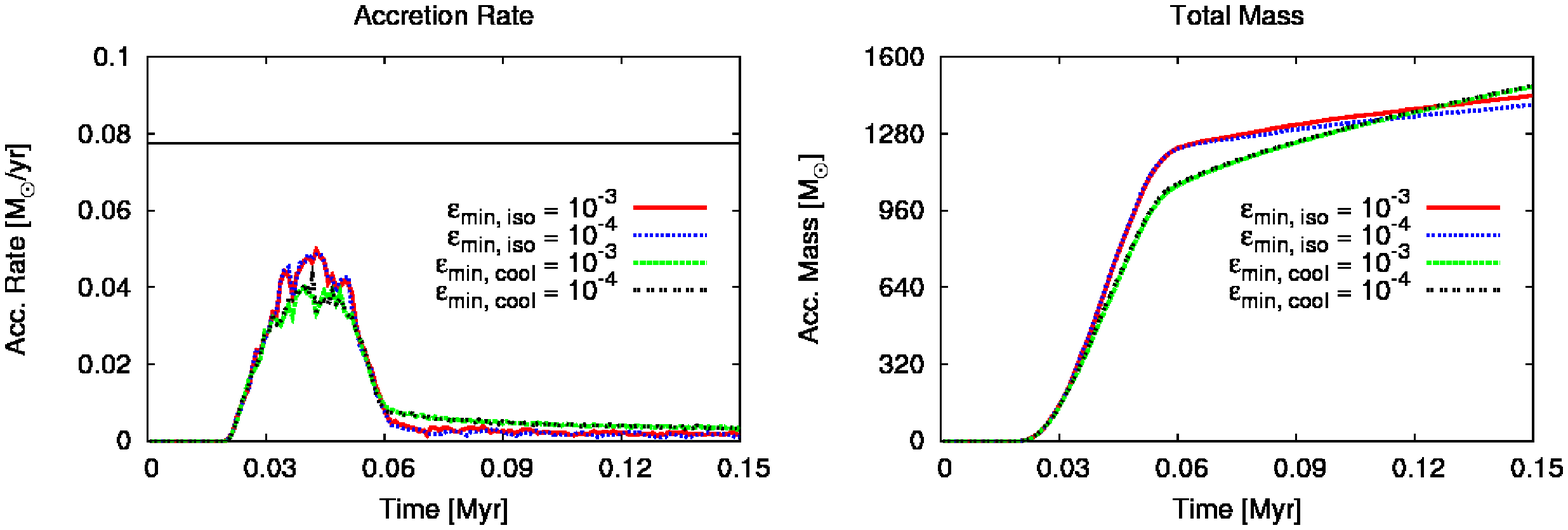}
\end{center}
\caption{
Accretion rate and total accreted mass as a function of softening length. Red (solid) shows
the results for $\epsilon = 10^{-3}$ pc, blue (dotted) the results for $\epsilon = 10^{-4}$ pc. 
To visualise the differences better the plots are shown only from 0 to 0.15 Myrs.
The difference is almost completely negligible.
\label{numstab_soft_a}}
\end{figure*}

\begin{figure*}
\begin{center}
\includegraphics[width=18cm]{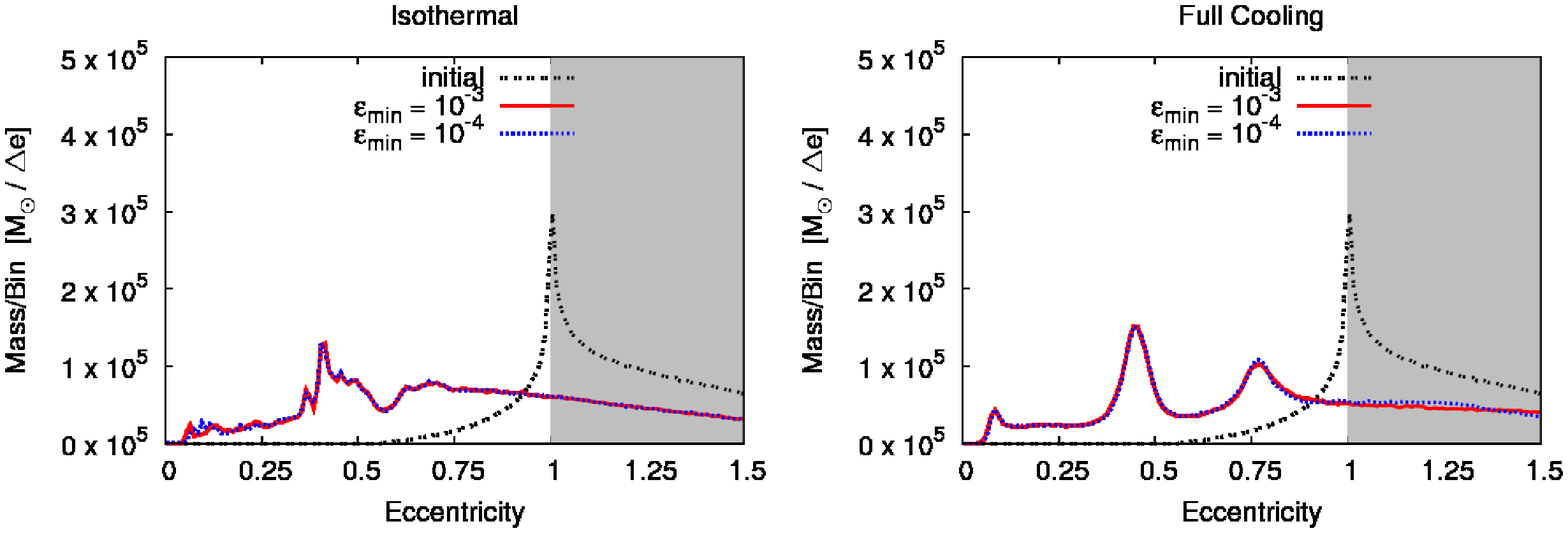}
\end{center}
\caption{Comparison of the distribution of mass onto orbits of different eccentricity depending on softening length.
Again the difference is almost completely negligible.
SPH particles with e $>$ 1 are not bound to the black hole, indicated by the grey background.
\label{numstab_soft_b}}
\end{figure*}

Finally, we studied the dependence of our results on the softening length. For this we 
repeated the full cooling and isothermal simulation using V03 with softening
length $\epsilon = 10^{-4}$ pc (compared
to our standard simulations with $\epsilon = 10^{-3}$ pc) making it very expensive in terms
of computing time. 
Figure \ref{numstab_soft_a} shows the resulting accretion rates and total accreted mass and
Figure \ref{numstab_soft_b} the distribution of mass onto orbits of different eccentricity. It can be clearly seen that the
difference in both cases is almost completely negligible. In the isothermal case we still do not resolve the
disc scale height to even one time the softening length, so that we expect the influence of an order
of magnitude smaller softening length
to have no real impact. In the cooling case we already resolve the disc scale height to 
multiple times the softening length for our standard value of $\epsilon = 10^{-3}$, thus the stability
of our results in this case indicates that we have reached convergence.


\end{appendix}
\label{lastpage}
\end{document}